\documentclass{article}

\usepackage[utf8]{inputenc}

\usepackage[english]{babel}

\usepackage[letterpaper,top=2cm,bottom=2cm,left=3cm,right=3cm,marginparwidth=1.75cm]{geometry}

\usepackage{amsmath,amssymb}
\usepackage{braket}
\usepackage{graphicx}
\usepackage{algorithm}
\usepackage{algpseudocode}
\usepackage[colorlinks=true, allcolors=blue]{hyperref}
\usepackage{multirow}
\usepackage[version=4]{mhchem}
\usepackage{authblk}
\usepackage[toc,page]{appendix}

\DeclareUnicodeCharacter{211D}{$\mathbb{R}$}

\title{Fully numerical Hartree-Fock calculations for atoms and small molecules with quantics tensor trains}
\author[a,b]{Paul Haubenwallner\footnote{\href{mailto:paul.haubenwallner@igd.fraunhofer.de}{paul.haubenwallner@igd.fraunhofer.de}}}
\author[a,b]{Matthias Heller\footnote{\href{mailto:matthias.heller@igd.fraunhofer.de}{matthias.heller@igd.fraunhofer.de}}}
\affil[a]{Fraunhofer Institute for Computer Graphics Research IGD, Darmstadt, Germany}
\affil[b]{Technical University of Darmstadt, Interactive Graphics Systems Group, Darmstadt, Germany
}
\date{}

\begin{document}
\maketitle
\vspace{-1cm}

\begin{abstract}
We present a fully numerical framework for the optimization of molecule-specific quantum chemical basis functions within the quantics tensor train format using a finite-difference scheme.
The optimization is driven by solving the Hartree-Fock equations (HF) with the density-matrix renormalization group (DMRG) algorithm  on Cartesian grids that are iteratively refined.
In contrast to the standard way of tackling the mean-field problem by expressing the molecular orbitals as linear combinations of atomic orbitals (LCAO) our method only requires as much basis functions as there are electrons within the system.
Benchmark calculations for atoms and molecules with up to ten electrons show excellent agreement with LCAO calculations with large basis sets supporting the validity of the tensor network approach.
Our work therefore offers a promising alternative to well-established HF-solvers and could pave the way to define highly accurate, fully numerical, molecule-adaptive basis sets, which, in the future, could lead to benefits for post-HF calculations.
\end{abstract}

\tableofcontents

\section{Introduction}
The Hartree-Fock (HF) method is one of the most basic tools in computer chemistry for simulating 
 the quantum mechanical behavior of molecules.
Traditionally, one performs HF calculations using a finite set of basis functions, which are linear combinations of so-called atomic orbitals~\cite{Jensen2017}.
In this approach, in order to accurately describe the molecular wave function, the number of the basis functions exceeds the number of electrons by a large factor, owing to the fact, that the basis functions are restricted in their expressiveness.
Various types of basis functions exist, including Slater-type orbitals (STO), Gaussian-type orbitals (GTO), and plane-wave basis functions, with GTOs being the most commonly used.
To achieve chemical accuracy within the LCAO approach even for relatively small molecules, hundreds or thousands of basis functions are needed, rendering calculations in the infinite-basis-size limit infeasible.

An interesting alternative to the LCAO-schemes is the finite-element-method (FEM) expressing wave functions through a set of polynomials defined in small overlapping volumes~\cite{White1989,Harrison2004,Lehtola2019}.
The main advantage of this method lies within the discretization of the space leading to local interactions of the basis functions as opposed to the LCAO approach in which all basis functions interact.
Although it has been argued that the FEM-calculations have a better scaling than the LCAO method, the overall number of the coefficients to get to chemical accuracy is quite high.

Going one step further down the representation ladder we arrive at fully numerical basis functions, i.e., functions which are defined on a grid with each grid point being independent on its neighbors.
The methodology of working with this kind of functions is dubbed finite-difference-method (FDM).
For correctly representing integrals important for quantum chemistry, even with advanced quadrature schemes an unmanageable amount of points is needed.
This makes the use of fully numerical basis functions for HF-calculations infeasible unless the simulation space is restricted or approximations are made.
A restriction of the simulation space can be achieved by only considering atoms or diatomic molecules, where it is possible to convert the three-dimensional problem to a two-dimensional one by using spatial symmetries of the system~\cite{Kobus2013}.

Starting with Ref.~\cite{Khoromskij2011a} there have been several successful implementations of approximating quantum chemical basis functions via tensor networks.
Here two different methodologies can be distinguished.
On one hand, in Ref.~\cite{Khoromskaia2018}, the authors compress the basis functions using the Tucker decomposition, but employ the standard LCAO approach.
The main benefit in these type of calculations stems from the fact that the tensorized orbitals allow for a fast and accurate calculation of the integrals while not necessarily being a well defined function like a Gaussian.
On the other hand, in Refs.~\cite{jolly2024, Marcati2022}, the authors discuss the direct optimization of the tensorized basis functions in real space.

For single-electron systems, this approach has been employed several times as numerical examples for a variety of tensor networks, using for example the DMRG algorithm.
For multi-electron systems, to the best of our knowledge, only Ref.~\cite{Rakhuba2016} discusses the solution of the mean-field problem by expressing the orbitals in the Tucker format.
A Block Green iteration is used to drive the solution of the HF-equations.
One notable example for post-HF calculations with tensor trains is given in Ref.~\cite{jolly2024} using an enrichment technique for optimizing the basis.

In this article, we present fully-numerical finite-difference HF-calculations based on tensor trains.
The molecular wave-function is expressed on a Cartesian three-dimensional lattice,  which, in order to avoid the curse of dimensionality, i.e., the exponential growth of data points needed to store the numerical values of the wave function, is expressed in terms of tensor trains.
Such a representation  has recently been introduced in the literature~\cite{jolly2024} and was also successfully used in numerical simulations in other contexts~\cite{Gourianov2022,Gourianov2025}.

The remainder of this article is structured as follows.
In Sec.~\ref{sec:intro_hf} we give a short introduction to the HF-method.
In Sec.~\ref{sec:tt_intro} we introduce the tensor train formalism and define the possible operations, that can be performed in this format.
In Sec.~\ref{sec:intro_quantics} we show how functional data defined on a Cartesian grid can be converted and manipulated within the tensor train formalism. 
In Sec.~\ref{sec:hf_tt} we combine the tools introduced in Sec.~\ref{sec:intro_quantics} and the Hartree-Fock formalism of Sec.~\ref{sec:intro_hf} for the construction of the HF-operators and subsequently the solution of the mean-field problem.
In Sec.~\ref{sec:results} we present some benchmark results of our method for molecules with a size of up to ten electrons.
In Sec.~\ref{sec:conclusion} we summarize our work and give an outlook.
In the Appendix we show some technical details on how to convert band matrices into the tensor train format {and how the difference of two wave functions can be quantified.}

\section{HF Method}
\label{sec:intro_hf}
The Hartree-Fock method (HF) is a standard method for solving the electronic-structure Schrödinger equation within the so called mean-field approximation, which neglects electron-electron correlations. For a pedagogical introduction see Ref.~\cite{Jensen2017}.
It is based on the ansatz that the $N_e$-electron wave function is expressed as a Slater determinant, which depends on single-electron wave functions $\phi_i(\mathbf{r}_i)$, the so called orbitals:
\begin{equation}
\Psi(\mathbf{r}_1, \mathbf{r}_2, \ldots, \mathbf{r}_{N_e}) = \frac{1}{\sqrt{N_e!}} \begin{vmatrix}
\phi_1(\mathbf{r}_1) & \phi_1(\mathbf{r}_2) & \cdots & \phi_1(\mathbf{r}_{N_e}) \\
\phi_2(\mathbf{r}_1) & \phi_2(\mathbf{r}_2) & \cdots & \phi_2(\mathbf{r}_{N_e}) \\
\vdots & \vdots & \ddots & \vdots \\
\phi_{N_e}(\mathbf{r}_1) & \phi_{N_e}(\mathbf{r}_2) & \cdots & \phi_{N_e}(\mathbf{r}_{N_e})
\end{vmatrix},
\label{eq:slater_determinant}
\end{equation}
where $N_e$ denotes the number of electrons and $\mathbf{r}_i$ denotes the position of the $i$-th electron.
The molecular Hamiltonian describes the interaction among the nuclei and the electrons of a molecule.
It is time independent and uses the so called Born-Oppenheimer approximation, in which only the electrons are treated quantum mechanically being trapped in the potential of the nuclei.
In atomic units it is given by
\begin{equation}
\begin{aligned}
    H_e &= T_e + V_{ne} + V_{ee} + V_{nn}, \\
    T_e &= \sum_{i=1}^{N_e} -\frac{1}{2} \Delta_i, \\
    V_{ne} &= \sum_{A=1}^{N_n} \sum_{i=1}^{N_e} \frac{-Z_A }{|\mathbf{r}_i - \mathbf{R}_A|},\\
    V_{ee} &= \sum_{i=1}^{N_e}\sum_{j>i}^{N_e} \frac{1}{|\mathbf{r}_i - \mathbf{r}_j|} \equiv \sum_{i=1}^{N_e} \sum^{N_e}_{j>1} g_{ij} ,\\
    V_{nn} &= \sum_{A=1}^{N_n}\sum_{B>A}^{N_n} \frac{Z_A Z_B}{|\mathbf{R}_A - \mathbf{R}_B|}.
    \label{eq:hf_operators}
\end{aligned}
\end{equation}
$H_e$ consists of the electronic kinetic energy $T_e$ and the three potentials $V_{ee}$, $V_{ne}$ and $V_{nn}$, expressing the electron-electron, the nuclei-electron and the nuclei-nuclei interaction, respectively.
$N_n$ denotes the number of nuclei, $\mathbf{R}$ and $Z$ their positions and charges.
For the following it is convenient to define the one-electron operator $h$ as
\begin{equation}
    h = - \frac{1}{2} \Delta + \sum_{A=1}^{N_n} \frac{-Z_A }{|\mathbf{r} - \mathbf{R}_A|}.\label{eq:kin_op}
\end{equation}
With the Coulomb and exchange operators $J_a$ and $K_a$,
{
\begin{equation}
\begin{aligned}
    J_a &= \int \frac{\phi_a(\mathbf{r}')\ \phi_a(\mathbf{r}')}{|\mathbf{r}-\mathbf{r}'|} d\mathbf{r}'  \\
    K_a\ket{\phi_b}&= \phi_a(\mathbf{r})\int \frac{\phi_a(\mathbf{r}')\phi_b({\mathbf{r}'})}{|\mathbf{r}-\mathbf{r}'|}d\mathbf{r}'\label{eq:coulomb_op}\\
\end{aligned}
\end{equation}
}
the energy of the ansatz is given by:
\begin{equation}
\begin{aligned}
    E(\phi_1,\dots \phi_n) = &\sum_{a=1}^{N_e}\bra{\phi_a}h\ket{\phi_a} + \frac{1}{2} \sum^{N_e}_{a=1,b=1} \bra{\phi_a}J_b\ket{\phi_a} - \bra{\phi_a}K_b\ket{\phi_a} + V_{nn}.
\end{aligned}
\label{eq:hf_energy}
\end{equation}

$E(\phi_1,\dots \phi_n)$ can be minimized using the variational principle and leads to the following set of pseudo-eigenvalue equations, also called Hartree-Fock (HF) equations:
\begin{equation}
F \phi'_a = \epsilon_a \phi'_a,
\label{eq:hf_equations}
\end{equation}
where $F$ is the Fock operator and $\epsilon_a$ the orbital energies.
The Fock operator is defined by
\begin{equation}
    F = h + \sum^{N_e}_{b=1} J_b - K_b.
    \label{eq:fock_operator}
\end{equation}
As can be seen the operator for minimizing the energy is dependent on the orbitals that are subject to the minimization.
The HF equations are therefore non-linear and can only be solved approximately using iterative methods.
Normally, for each iteration the operators are calculated based on an initial guess or the results of the last iteration and used to solve Eq.~\eqref{eq:hf_equations}.
This process is repeated until convergence criteria are met.

Electrons are fermions and thus obey the Pauli principle, which states, that two particles in a system cannot have the same quantum numbers.
Due to this, the HF calculations can be simplified such, that two electrons are placed in a single orbital with different spin values.
This effectively reduces the number of orbitals by a factor of two.
The method is then dubbed restricted Hartree-Fock.

\section{TTs and TT matrices}
\label{sec:tt_intro}
\begin{figure}[b!]
    \centering
    \includegraphics[width=0.75\linewidth]{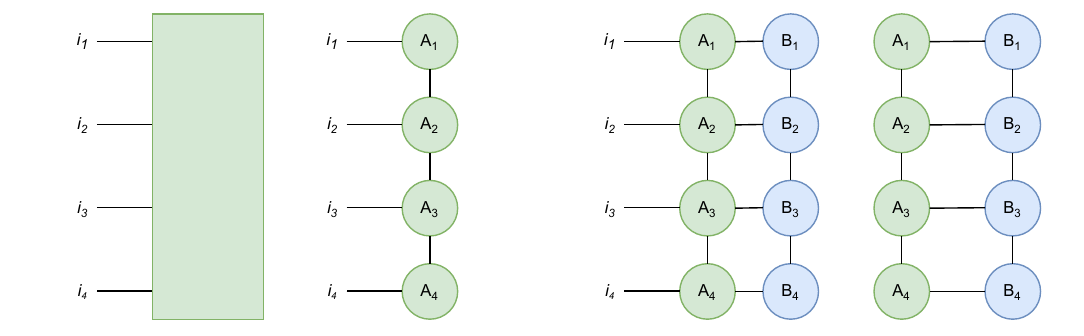}
    \caption{Pictorial representation of a general tensor and a tensor train (left panel) and the two tensor train operations ${\rm TTM}_A \cdot {\rm TT}_B$ and ${\rm TT}_A \cdot {\rm TT}_B$ (right panel). Open lines denote indices of the tensor, closed lines denote a contracted pair of indices.\label{fig:tensor_diagrams}}
\end{figure}

Tensor trains (TT), also called marix-product states (MPS), are linear tensor networks, which have been used extensively to describe one-dimensional quantum systems in the past --- for a recent review see Ref.~\cite{Dey2023}.
Given an arbitrary tensor $T^{i_1\cdots i_n}$ with $n$ open indices of dimension $d$, the tensor train decomposition of $T$ is given by
\begin{equation}
    T^{i_1\cdots i_n} \approx  A^{i_1}_{r_1}\cdot A^{i_2}_{r_1r_2} \cdot\ \dots\ \cdot A^{i_{n-1}}_{r_{n-1}r_n} \cdot A^{i_n}_{r_n} \equiv \prod^n_{a=1} A^{i_a}_a,
    \label{eq:tensor_train}
\end{equation}
where the tensors $A$ are called cores and have the shape $(d,r_k,r_{k+1})$ with the so called core ranks, or also called bond dimensions, $r_k$ and $r_{k+1}$.

As is evident from Eq.~\eqref{eq:tensor_train}, the tensor train decomposition of a tensor has in total $\sum_i d r_i r_{i+1}$ parameters, while the tensor on the lhs.~of Eq.~\eqref{eq:tensor_train} has in total $d^n$ parameters.
The efficiency of the tensor train representation becomes particularly useful when dealing with tensors described by many external indices. 
This is because the total number of parameters in the original tensor $T^{i_1\cdots i_n}$ increases exponentially while the tensor train decomposition allows for a more compact representation, especially when the bond dimensions $r_i$ can be kept small.

Tensors and more specific tensor trains, can be easily visualized through simple diagrams.
In these diagrams, a line represents an index, and when two lines are connected, it indicates the contraction of both indices.
Tensors are usually depicted as boxes or circles.
An example of a tensor as well as a tensor train with $4$ dimensions, i.e., with $4$ open indices, is shown in the left panel of Fig.~\ref{fig:tensor_diagrams}.
Note, that the matrix product between the different tensor cores (depicted as circles) in the tensor train as defined in Eq.~\eqref{eq:tensor_train} are depicted through connected lines.

Not every tensor admits a low-rank tensor train decomposition. In the context of simulating quantum mechanical systems, it is well-established that the eigenstates of one-dimensional, local Hamiltonians can be described efficiently by low-rank tensor trains, which can be explained by the area-law of entanglement~\cite{Hastings2007}.
In this work, we use tensor trains to describe exponentially many data points of three-dimensional functions on a Cartesian grid.
Here, a tensor network representation with low ranks exists if the function exhibits internal structure in the sense of scale separation~\cite{Ye2022}.

In addition to tensor trains, we also need the concept of tensor train matrices (TTM), also called matrix product operators in the literature. 
Given a tensor with $2n$ external indices $(i_1,j_1,\dotsc i_n,j_n)$, the decomposition is given by
\begin{equation}
    T^{i_1j_1\cdots i_nj_n} \approx A^{i_1j_1}_{r_1}\cdot A^{i_2j_2}_{r_1r_2} \cdot\ \dots\ \cdot A^{i_{n-1}j_{n-1}}_{r_{n-1}r_n} \cdot A^{i_nj_n}_{r_n} \equiv \prod^n_{a=1} A^{i_aj_a}_a,
    \label{eq:tensor_train_op}
\end{equation}
where the cores $A^{i_k j_k}_k$ are now tensors of shape $(d,d,r_k,r_{k+1})$.
In quantum physics tensor train matrices are typically used to encode the Hamiltonian or any other Hermitian operator.

There is a natural equivalence between tensor trains (tensor train matrices) and vectors (matrices).
In the same way as one can define matrix-vector multiplication, we define the multiplication between a tensor train matrix and a tensor train. This is achieved by contracting over the set of open indices shared by the tensor train and the tensor train matrix:
\begin{equation}
    {\rm TTM} \cdot {\rm TT} = \sum_{(j_1,\dots, j_n)} \left(A_1^{i_1 j_1}\cdot A_2^{i_2 j_2} \cdots A^{i_n j_n}_n\right)  \left(B_1^{j_1}\cdot B_2^{j_2} \cdots B^{j_n}_n\right).
\end{equation}
This operation yields a new tensor train, given by
\begin{equation}
    {\rm TTM} \cdot {\rm TT} =C_1^{i_1}\cdot C_2^{i_2} \cdots C^{i_n}_n,
\end{equation}
where the cores are defined by $C_k^{i_k} = \sum_{j_k} A_k^{i_k j_k}  B_k^{j_k}$ and have tensor rank $r_a\cdot r_b$
The diagrammatic representation of a multiplication of a tensor train matrix with a tensor train is shown in the right panel of Fig.~\ref{fig:tensor_diagrams}.

\begin{table}[b!]
    \centering
    \begin{tabular}{c|c}
        operation & resulting bond dimension \\ \hline 
        $r \cdot {\rm TT}$ & $r$ \\
        ${\rm TTM}_a \cdot {\rm TT}_b$ & $r_a \cdot r_b$ \\
        ${\rm TT}_a + {\rm TT}_b$  & $r_a+r_b$  \\
        ${\rm TT}_a \cdot {\rm TT}_b$  & $1$  
    \end{tabular}
    \caption{Arithmetic operations defined for tensor trains and their effect on the bond dimension. The last operation (dot product between TTs) results in a scalar.}
    \label{tab:bond_dim_ops}
\end{table}
In addition to the matrix-vector product one can also define other operations for tensor trains, such as summation, multiplication by a scalar or the dot product of two tensor trains, which results in a scalar. 
In order to perform a sum of two tensor trains with cores $A_k$ and $B_k$, one can construct a new tensor train with cores $C_k$ defined by
\begin{equation}
    C_k = \left( \begin{array}{cc}
       A_k  & 0 \\
       0  &  B_k
    \end{array}\right),
\end{equation}
such that the bond dimension of $C_k$ is evidently given by ${\rm dim}(A_k)+{\rm dim}(B_k)$.
For multiplication of a tensor train with a scalar $x$, one can choose one core (for example the first one) and rescale it appropriately: $A_0 \rightarrow x A_0$.
The dot product of two tensor trains can be performed as
\begin{equation}
    {\rm TT}_a \cdot {\rm TT}_b = \sum_{(j_1,\dots, j_n)} \left(A_1^{j_1}\cdot A_2^{j_2} \cdots A^{j_n}_n\right) \left(B_1^{j_1}\cdot B_2^{j_2} \cdots B^{j_n}_n\right),
\end{equation}
which after summation over all indices $j_k$ results in a scalar value.
All operations and how they effect the bond dimension are summarized in Tab.~\ref{tab:bond_dim_ops}.

In practice one has to make sure that the tensor cores do not become too large, such that computations are still feasible.
There are multiple algorithms that are able to perform this kind of compression.
A detailed overview of these can be found in Ref.~\cite{Paeckel2019}.

\section{Quantics TT representation of the finite-difference method}
\label{sec:intro_quantics}

\begin{figure}
    \centering
    \includegraphics[width=1\linewidth]{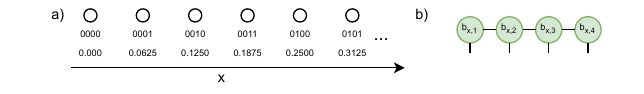}
    \caption{a) Example for the discretization of a variable x with a binary fraction according to~\eqref{eq:binary_fraction} with $x \in [0,1)$ and $p_d=4$. Displayed is the evenly spaced grid, the bitstring representation of the value as well as the value itself for the first six grid points. b) Corresponding tensor train in the quantics representation.}
    \label{fig:binary_fraction}
\end{figure}

The so called quantics tensor train (QTT) representation \cite{Oseledets2009, Khoromskij2011} describes the approximation of an arbitrary function $f(x_1,x_2,...,x_n)$ via tensor trains with exponential precision.
It is based on binary fractions, where each variable $x_i \in [x_{i,\text{min}}, x_{i,\text{max}})$ is discretized up to a precision of $p_i$ such that
\begin{equation}
    x_i(b_{x_i,1}, b_{x_i,2}, ..., b_{x_i,p_i}) = \sum^{p_i}_{j=1} \frac{b_{x_i,j}}{2^j}\ (x_{i,\text{max}} - x_{i,\text{min}}) + x_{i,\text{min}},
    \label{eq:binary_fraction}
\end{equation}
with $b_{x_i,j} \in \{0, 1\}$ being the bit variable.
In this way $x_i$ is mapped to the bit values $b_{x_i,j}$ and can be described as a bitstring with $p_i$ bits (see Fig.~\ref{fig:binary_fraction} a).
Another way of thinking about the binary fractions is that each variable is defined by an evenly spaced one dimensional grid with $2^{p_i}$ values in the range $[x_{i,\text{min}}, x_{i,\text{max}})$.
So with every additional bit of precision the number of the values doubles.
The quantics representation assigns each bit to one open index in the tensor train (see Fig.~\ref{fig:binary_fraction} b), which therefore has a length of $\sum^n_i p_i$.
Due to a bit only having two possible values, the open indices have dimension two.

\subsection{Encoding} \label{sec:Encoding}
The encoding in the QTT format defines the contraction order of a tensor train, i.e., the sequence in which the tensor cores with the open indices of the binary fractions ($b_{x_i,j}$) are arranged.
It has a big impact on the quality of the approximation and cannot be changed easily during the computation.
For an efficient application of tensor train matrices to tensor trains the encodings of both have to match.

\begin{figure}[b!]
    \centering
    \includegraphics[width=1.00\linewidth]{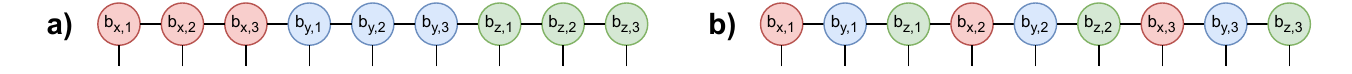}
    \caption{a) Block encoding and b) alternating encoding for a function $f(x,y,z)$ with $p_x=p_y=p_z=3$.}
    \label{fig:encoding}
\end{figure}

Since a reversed order leads to the same encoding, the total number of all unique encodings with $N$ bits is given by $\frac{N!}{2}$.
It is not a priori known, which encoding leads to the best approximation of a function, so the ordering is normally established on the basis of certain rules.
The two most-commonly used encoding schemes, that can be found in the literature, are the block encoding and the alternating encoding \cite{Marcati2022, Kazeev2017} (see Fig.~\ref{fig:encoding}).

In the block encoding the bits of each variable $x_i$ are grouped together in a block, where the order in each block is arranged from the largest to the smallest fraction.
In the alternating encoding the bits describing the same fraction are grouped together and the corresponding groups are sorted in ascending order.
The original formulation also contracts each bit-group (for example $x_1$-$y_1$-$z_1$) leading to tensor trains with multiple open indices at each core~\cite{Kazeev2017}.

\subsection{Tensor Cross Interpolation (TCI)}
Tensor cross interpolation (TCI) algorithms \cite{Oseledets2011, Oseledets2010, Savostyanov2011, Savostyanov2014, Dolgov2020} allow the efficient transformation of high dimensional black-box functions into tensor trains.
Within the QTT picture the naive approach of compressing a discretized function into a tensor train involves the evaluation of the function at all grid points followed by multiple higher-order singular value decompositions (HOSVD) (for an excellent introduction see Ref.~\cite{NunezFernandez2022}).
This approach is considered the best for achieving low and controllable approximation errors and core ranks but comes with the price of an exponential scaling.
It can therefore only be employed for small systems.
TCI algorithms on the other hand are capable of approximating functions with much less function calls albeit at the price of larger core ranks but still with controllable errors.
This makes them ideal for compressing functions defined on exponentially fine grids that can be efficiently approximated within the QTT representation \cite{Ritter2024}.

Besides tensor trains it is also possible to approximate tensor train matrices.
Since the TCI algorithms are not dependent on a certain dimensionality of the open indices, the operators can be approximated directly.
Another approach is to rewrite the tensor train matrix as a tensor train by splitting each TTM-core into two TT-cores.
After learning the tensor train each set of the split TTM-cores can be contracted to form again the tensor train matrix.

\subsection{Band Operators}
\label{sec:band_op}
\begin{figure}[b!]
    \centering
    \includegraphics[width=1.0\linewidth]{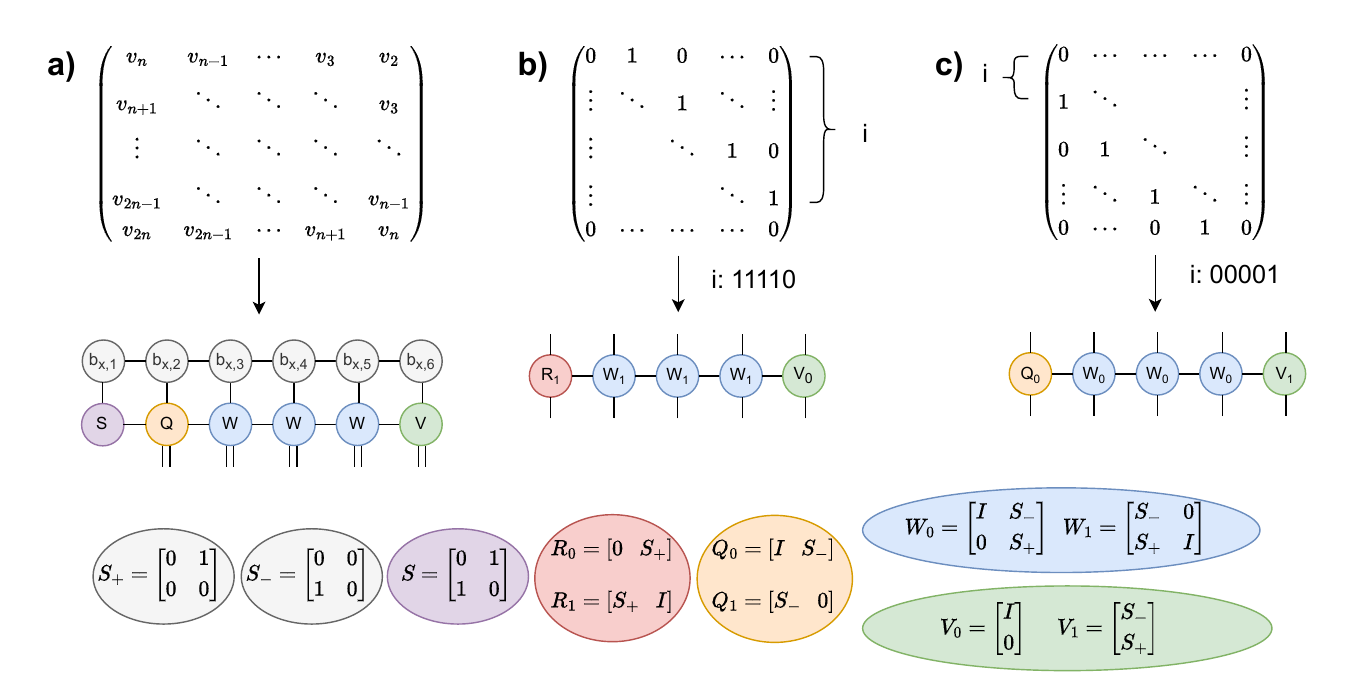}
    \caption{QTT-representation of different band matrices for a single dimension. a) depicts the conversion of a  tensor train of length $l+1$ to a Toeplitz TTM of length $l$. b) and c) depict how to construct tensor train matrix with rank-2 for a single band dependent on their distance to the main diagonal. For obtaining multilevel band matrices one can apply the same transformations for each sequence of cores describing the same dimension. {For a in-depth description of the operations we refer to~\cite{Kazeev2013}}.}
    \label{fig:band_mat}
\end{figure}

Performing simulations with tensor trains can lead to a significant increase in performance, so that classically intractable systems can be solved.
This is however only the case if tensor trains as well as the corresponding operators, the tensor train matrices, represent a good approximation of the problem.

An important type of matrices, that are used to represent several finite-difference operators, are the so called band matrices.
A band matrix is a sparse matrix with equal, non-zero elements that are confined to one or multiple diagonal bands, like for instance
\begin{equation}
    \begin{pmatrix}
    2 & 0 & 3 & 0\\
    1 & 2 & 0 & 3\\
    0 & 1 & 2 & 0\\
    0 & 0 & 1 & 2\\
    \end{pmatrix}.
\end{equation}
The tensor train matrix for the main diagonal has rank-1 where each core is an identity matrix.
For all other diagonals the structure of the TTM is less trivial.
Appendix~\ref{sec:appendix_a} introduces a method for representing each diagonal band as a tensor train matrix for arbitrary encodings.
Since a band matrix can be decomposed into a sum of single-band matrices, it is therefore possible to transfer any band matrix into a TTM.
As can be seen in Fig.~\ref{fig:band_ops_algo} the approach is rather inefficient, but illustrates the process of finding a single-band matrix quite well.

If one uses the block encoding, more efficient ways of constructing band matrices as TTMs have already been reported in the literature~\cite{Kazeev2013}.
In that case, it has been proven that single-band matrices can be exactly described by rank-$2$ TTMs.
In addition to that, arbitrary tensor trains can be converted to different kinds of Toeplitz matrices while doubling the rank.
This includes the general and circulant Toeplitz matrix as well as the upper and lower triangular ones.
Note, that the transformations are valid for multiple dimension.
Fig.~\ref{fig:band_mat} depicts the construction of single-band matrices as well as the conversion of a tensor train into the general Toeplitz matrix.

\subsubsection{Laplace Operator} \label{sec:LaplaceOperator}
The Laplace operator, most commonly denoted by $\Delta$, is the second-order differential operator describing the divergence of the gradient.
{It is used in the formulation of many different eigenvalue problems and also occurs in the HF equations.}
In the finite-difference-method, $\Delta$ is calculated numerically by considering the difference of neighboring points divided by the grid spacing.
In this work, we use central finite differences, for which one considers left- and right-lying neighbors at a given point.
For instance, to lowest accuracy, the second derivative of a function $f(x)$ at grid point $x_n$ is given by
\begin{equation}
    f''(x_n) = \frac{f(x_{n-1})-2f(x_{n})+f(x_{n+1})}{h^2},\label{eq:finite_diff_example}
\end{equation}
where $h$ is the space between two grid points.
Higher-order approximations of the Laplace operator can be calculated, by interpolating between more neighboring points.
The coefficients for an accuracy up to eighth order can be found in Ref.~\cite{Fornberg1988} and are shown in Tab.~\ref{tab:finite_diff_coeffs}.
\begin{table}[b!]
    \centering
    \begin{tabular}{|c|c|c|c|c|c|c|c|c|c|}\hline
        Accuracy & \multicolumn{9}{c|}{Coefficients}  \\ \hline
            & $-4$ & $-3$ & $-2$ & $-1$ & $0$ & $1$ & $2$ & $3$ & $4$ \\ \hline
        $2$ & & & & $1$ & $-2$ & $1$ & & & \\
        $4$ & & & $-1/12$ & $4/3$ & $-5/2$ & $4/3$ & $-1/12$ & & \\
        $6$ & & $1/90$ & $-3/20$ & $3/2$ & $-49/18$ & $3/2$ & $-3/20$ & $1/90$ & \\
        $8$ & $-1/560$ & $8/315$ & $-1/5$ & $8/5$ & $-205/72$ & $8/5$ & $-1/5$ & $8/315$ & $-1/560$ \\ \hline
    \end{tabular}
    \caption{Coefficients to calculate the second derivative of a function using the finite difference method up to an accuracy of order $8$.}
    \label{tab:finite_diff_coeffs}
\end{table}

In order to derive a tensor train matrix that performs the operation of Eq.~\eqref{eq:finite_diff_example}, note that we can rewrite this for all points on the grid using a band matrix:
\begin{equation}
\begin{pmatrix}
    f''(x_0)\\
    f''(x_1)\\
    \cdots\\
    \cdots\\
    f''(x_n)\\
    \end{pmatrix}=
     \begin{pmatrix}
    -2 & 1 & 0 & \cdots & 0\\
    1 & -2 & 1  & \cdots & 0\\
    0 &1 & -2 &  \cdots & 0\\
    \cdots &  \cdots &  \cdots &  \cdots &  \cdots\\
    0 &0 &  \dots & \cdots & -2\\
    \end{pmatrix}\cdot \begin{pmatrix}
    f(x_0)\\
    f(x_1)\\
    \cdots\\
    \cdots\\
    f(x_n)\\
    \end{pmatrix}.
\end{equation}

Evidently, we can write the Laplace operator as a sum of three different band operators with diagonal values $-2$ and $1$.
Using the algorithms introduced in the previous section we can then derive the corresponding tensor train matrix.
The same method also works for higher-order approximations of the Laplace operators, when taking into account more neighboring points.

\subsubsection{Displacement Operator} \label{sec:Displacement}
\begin{figure}
    \centering
    \includegraphics[width=0.75\linewidth]{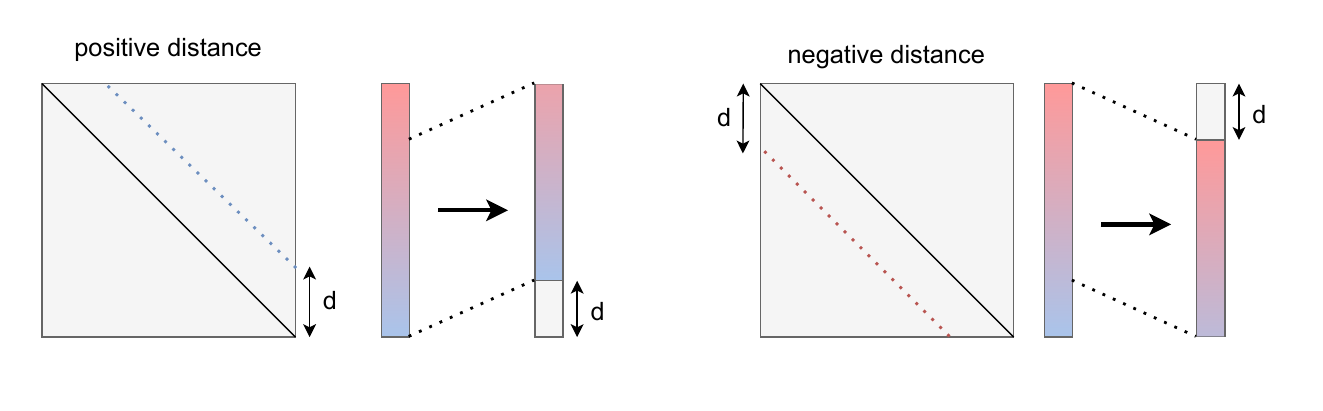}
    \caption{Shifting operation of a single band matrix for positive and negative distances (in respect to the main diagonal).}
    \label{fig:band_operator_shifting}
\end{figure}
The displacement of a discretized function {(for example $\frac{1}{\mathbf{|r|}}$ by $\mathbf{R}$, see \eqref{eq:hf_operators})} can be realized by shifting each grid value by a specified distance.
This is nothing else than applying a single-band matrix to a vector (see Fig.~\ref{fig:band_operator_shifting}) and is therefore an efficient operation in the QTT representation.
In one dimension the displaced tensor train is
\begin{equation}
    \text{TT}_{\text{disp}} = B(d)\ \text{TT},
\end{equation}
where the tensor train matrix $B(d)$ approximates a single-band operator and is dependent on the distance $d$.
If the tensor train is based on a multi-dimensional function $f(x_1, x_2, ..., x_n)$ the operator $B(d_{x_1},d_{x_2},...,d_{x_n})$ can be decomposed into a product of band operators for each dimension
\begin{equation}
    \text{TT}_{\text{disp}} = \prod^n_i B_{x_i}(d_{x_i})\ \text{TT}.
\end{equation}
The function can be displaced by either calculating $\prod^n_i B_{x_i}(d_{x_i})$ before applying it to the tensor train or by applying each $B_{x_i}(d_{x_i})$ successively.
Note that a displacement operation as described above leads to a loss of information, since the values at the boundaries of the grid are either set to zero or moved outside the domain.

\subsection{Interpolation} \label{sec:Interpolation}
Interpolation is the approximation of new data points in a certain range based on existing data points within this range.
{In FDM simulations it can be used to switch between varying fine grids.}
In the QTT representation, the interpolation of a tensor train increases its length by one, doubling the data points of the corresponding grid.
The added points need to be approximated by the courser grid.

\begin{figure}
    \centering
    \includegraphics[width=0.75\linewidth]{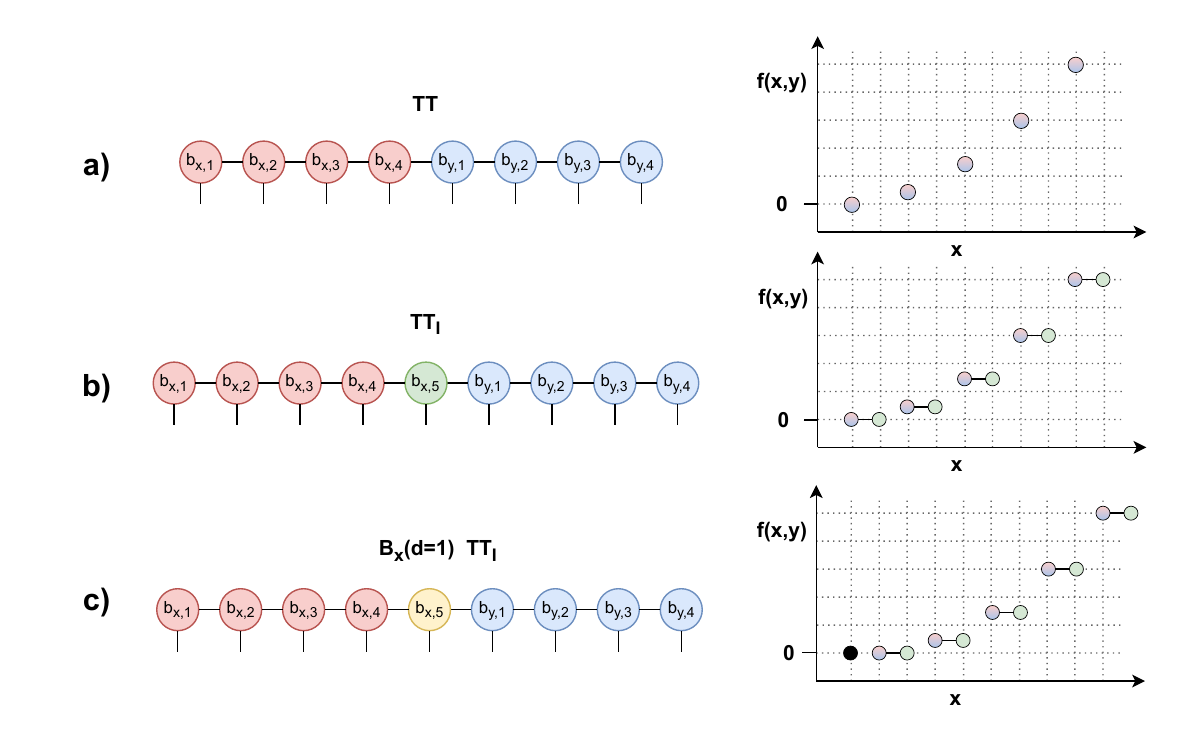}
    \caption{{Graphical depiction of tensor trains needed for the interpolation process for an two dimensional example in block encoding (left) and their corresponding functional values (right) according to \eqref{eq:tt_interpolation}. a) shows the original tensor train that needs to be interpolated. b) shows the tensor train, where an identity tensor is added along the x-axis. c) shows the same operator with a shift by one applied to it.}}
    \label{fig:interpolation}
\end{figure}

Following \cite{GarciaRipoll2021} a linear interpolation for one dimension can be efficiently performed.
Here each interpolated point $f(x_1,x_2,...,x_{i,j},x_n)$ of dimension $i$ at grid position $j$ is calculated by the neighbors and can be expressed as
\begin{equation}
    f(x_1,x_2,...,x_{i,j},...,x_n) = \frac{f(x_1,x_2,...,x_{i,j-1},...,x_n)+f(x_1,x_2,...,x_{i,j+1},...,x_n)}{2}.
    \label{eq:linear_interpolation}
\end{equation}
To perform an equivalent operation on a QTT grid, the first step is to increase the length of the tensor train ({denoted as $\text{TT}$ in Fig.~\ref{fig:interpolation} a)}) by appending a core, which describes the added values on the finer grid.
By initializing the new core with identity matrices for the open indices 0 and 1, the added values are equal to the old values and a new tensor train is created ({denoted as $\text{TT}_I$ in Fig.~\ref{fig:interpolation} b)}).
For the interpolation a second tensor train is needed.
It is defined by $\text{TT}_I$ shifted by the displacement operator $B_{x_i}(d_{x_i}=1)$ ({Fig.~\ref{fig:interpolation} c)}).
The interpolated tensor train can now be described as
\begin{equation}
    \text{TT}_{\text{intpl.}} = \frac{\text{TT}_I+B_{x_i}(d_{x_i}=1)\text{TT}_I}{2}.
    \label{eq:tt_interpolation}
\end{equation}
It should be mentioned that the position of the introduced core is completely arbitrary and does not necessarily have to obey any predefined ordering.
That being said, normally the position is based on the aforementioned encoding rules.

The opposite of linear interpolation, the mapping of a function from a finer grid to a courser grid, is also possible.
It can be simply achieved by contracting the core describing the smallest fraction of a dimension with one of the neighboring cores.

\subsection{Adding dummy Indices} \label{sec:AddingDimensions}
Adding dummy indices to a tensor train or a tensor train matrix is a useful tool to describe operations in the QTT representation.
An additional index can be simply added to a tensor train with the cores $A^a_{ij}$ by performing
\begin{equation}
    A^{aa'}_{ij} = A^a_{ij} \cdot I^{aa'}
\end{equation}
for each core separately, where $I^{aa'}$ is a $2 \times 2$ identity matrix.
This operation does nothing but the transformation of the tensor train into a diagonal tensor train matrix.
By doing so it is possible to describe the multiplication of two tensor trains diagramatically as the contraction of a diagonal tensor train matrix and a tensor train.
The same can be also done for tensor train matrices with the cores $A^{ab}_{ij}$ 
\begin{equation}
    A^{aa'b}_{ij} = A^{ab}_{ij} \cdot I^{aa'}.
\end{equation}

\section{Solving the HF equation with TTs}
\label{sec:hf_tt}
In this section we are concerned with the solution of the restricted Hartree-Fock equations in real space by approximating the finite-difference method with tensor trains.
Within a three-dimensional Cartesian coordinate system $(x, y, z)$, the one-electron wave functions $\phi_a(\mathbf{r}_i)$ and operators like $g(\mathbf{r}_i, \mathbf{r}_j)$ are functions of the electrons coordinate vector
\begin{equation}
    \mathbf{r}_i = \begin{pmatrix} x_i \\ y_i \\ z_i \end{pmatrix}.
\end{equation}
The task is therefore to express functions $f(x_i,y_i,z_i)$ and operators $o(x_i,y_i,z_i,x_j,y_j,z_j)$ as tensor trains and tensor train matrices, respectively.
As described above this can be done with the QTT representation on an evenly spaced grid.
The formulations derived and used in the following do in principle not depend on the encoding of the tensor trains.
However, as mentioned in Sec.~\ref{sec:band_op}, some operations, like the transformation of a tensor train into a Toeplitz matrix, have a straight-forward formulation only within the block encoding.
It is therefore more convenient to express the following algorithms within the block encoding, in which all dimensions are separated.

Having established the basic algorithms, first it is necessary to describe how the functions and operators defining the HF method can be converted.
For an overview see Fig.~\ref{fig:operatorsl}.
After that, the tensor train adjusted minimization routine (Fig.~\ref{fig:minimization}) is illustrated.

\subsection{Orbitals and Fock Operators}
\label{sec:fock_ops}
Since the Fock operators are partially dependent on the orbitals, every method based on Eq.~\eqref{eq:hf_equations} needs a starting guess for the orbitals $\phi(x,y,z)$.
Assuming $\phi(x,y,z)$ can be efficiently sampled, the transformation of an orbital into a quantics tensor train can be simply done by tensor cross interpolation.
The assumption is justified, if the initial guess is based on standard Gaussian- or Slater-type orbitals.

\begin{figure}[b!]
    \centering
    \includegraphics[width=1\linewidth]{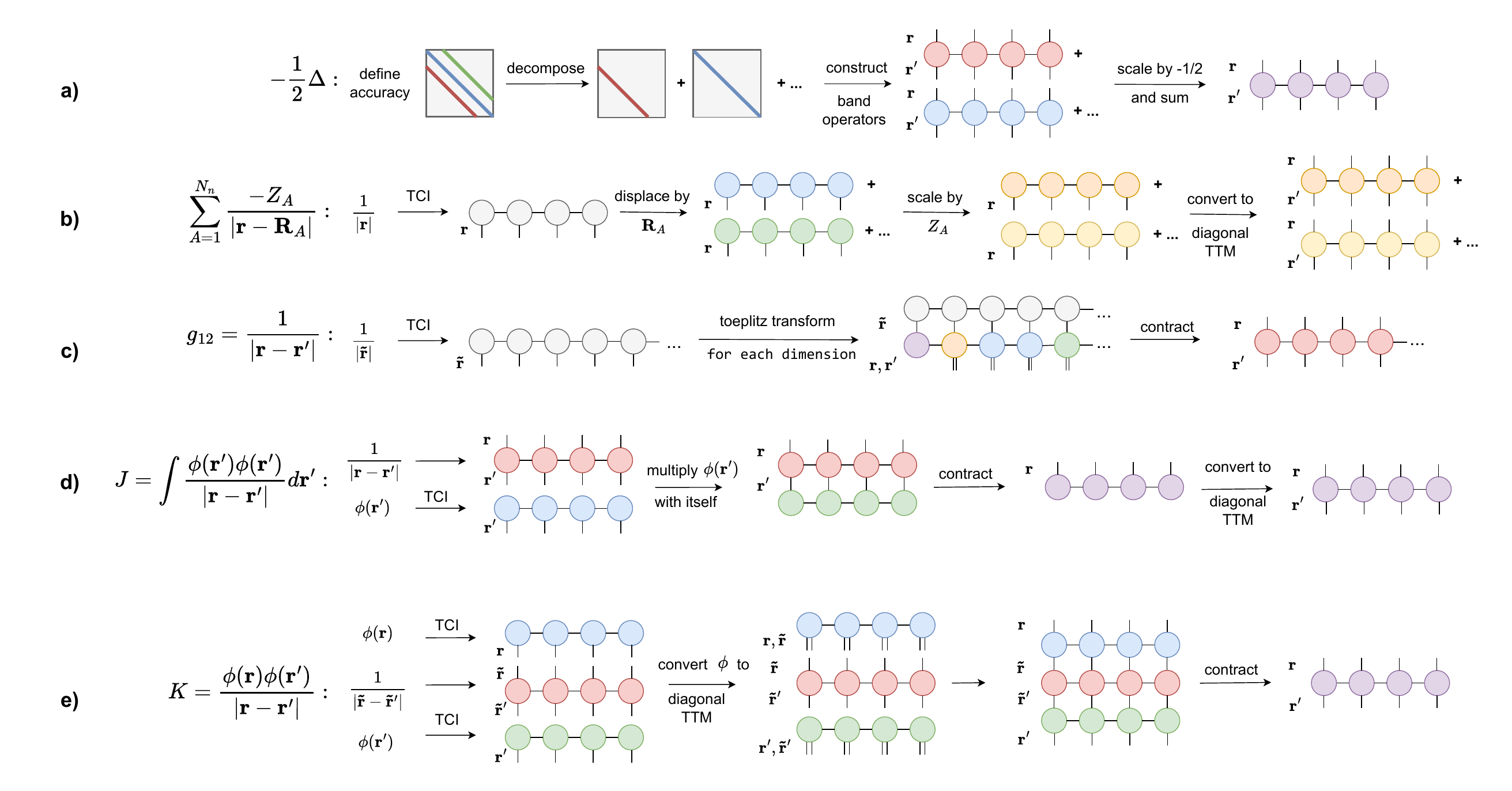}
    \caption{Illustration of constructing the Fock operators in the QTT representation. TCI stands for tensor cross interpolation and converts the functions to tensor trains or tensor train matrices. The displacement is done according to \ref{sec:Displacement}. {Tensor trains are converted to diagonal TTMs} following the steps in \ref{sec:AddingDimensions}. The Toeplitz transformation (for one dimension) of a tensor train can be seen in Fig.~\ref{fig:band_mat}.}
    \label{fig:operatorsl}
\end{figure}

Turning to the actual operators, the kinetic energy of the electronic wave function is defined by the Laplace operator acting on each orbital separately, multiplied by a scalar.
Because of this, only a single tensor train matrix describing $-\frac{\Delta}{2}$ is needed.
Following the results of section \ref{sec:LaplaceOperator} a TTM with low ranks can be constructed via band operators.
{
The process for this is depicted in Fig~\ref{fig:operatorsl} a).
After choosing the accuracy the distinct bands of the matrix can be represented by band operators of rank-2.
$\Delta$ is the sum of these operators scaled by a factor of $-1/2$.
}

The next operator to be expressed as a tensor train matrix is the electron-nuclei potential $V_{ne}$.
In Cartesian coordinates, $V_{ne}$ can be written for one electron coordinate $\mathbf{r}=(x,y,z)$ as 
\begin{equation}
    V_{ne} = \sum^{N_n}_{A=1}\frac{-Z_A}{\sqrt{(x-X_A)^2 + (y-Y_A)^2 + (z-Z_A)^2}}.
    \label{eq:pot_ec}
\end{equation}
The potential is the sum of multiple inverse distance operators centered around the nuclei and scaled by their charge.
Thus it is sufficient to perform a cross interpolation for $\frac{1}{|r|}$ once with a subsequent displacement of the tensor train to the corresponding nuclei positions by using band operators ({first two steps in Fig.~{\ref{fig:operatorsl}}} b)).
The resulting tensor train describes $\frac{1}{|\mathbf{r}-\mathbf{R}_A|}$ and just needs to be multiplied with a scalar to represent one single term of Eq.~\eqref{eq:pot_ec}.
To obtain the operator $V_{ne}$ the tensor trains are converted to diagonal tensor train matrices.

Having depicted how the one-electron operators can be transformed, we now turn towards the two-electron operators $J$ and $K$.
The central element that both of these operators have in common is the inverse distance operator dependent on two electron coordinates $g_{12}$.
In Cartesian coordinates it is expressed by 
\begin{equation}
    g_{12} = \frac{1}{\sqrt{(x_1-x_2)^2 + (y_1-y_2)^2 + (z_1-z_2)^2}}.
\end{equation}

In contrast to the $\frac{1}{|r|}$ operator $g_{12}$ is not a diagonal matrix but a multilevel Toeplitz matrix.
It can be approximated by learning $\frac{1}{|r|}$ with one additional core in each dimension using a cross interpolation followed by a multilevel Toeplitz transformation (see Fig.~\ref{fig:operatorsl} c)).

With $g_{12}$ given as a tensor train matrix and $\phi(\mathbf{r})$ as tensor train, $J$ and $K$ can be constructed by the appropriate contractions for each orbital.
For calculating $K$ one has to turn $\phi(\mathbf{r})$ to a diagonal TTM and multiply it from both sites to $g_{12}$ {as depicted in Fig.~\ref{fig:operatorsl} e)}.
The operations for $J$ are slightly different.
{Here one multiplies $\phi(\mathbf{r}')$ with itself to gain $\phi(\mathbf{r}')^2$ and subsequently contracts $\phi(\mathbf{r}')^2$ with $g_{12}$ from one side (Fig.~\ref{fig:operatorsl} d)), effectively integrating over one electron coordinate.
The result is a tensor train and can be easily converted to a diagonal TTM completing the construction of $J$.}

To summarize we have the non-diagonal kinetic operator, a sum of diagonal operators for the electron-nuclei potential as well as the diagonal Coulomb operators and the non-diagonal exchange operators.
According to Eq.~\eqref{eq:fock_operator} this is sufficient for all Fock operators needed to perform a Hartree-Fock calculation.

\subsection{Minimization Routine}
\begin{figure}[b!]
    \centering
    \includegraphics[width=0.85\linewidth]{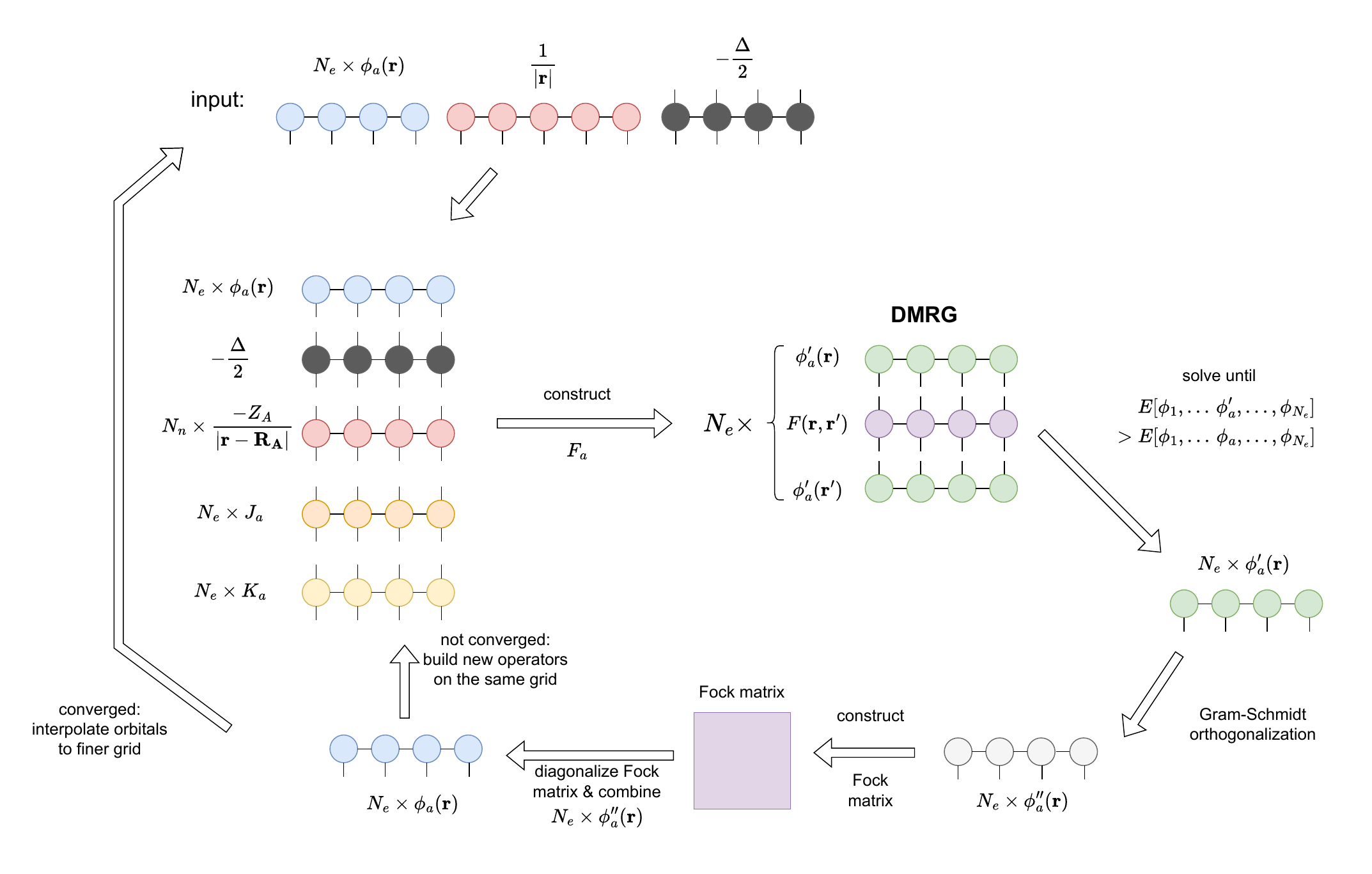}
    \caption{Minimization routine for performing Hartree-Fock calculations within the QTT representation. {The input is an initial set of orbitals $\phi$, a tensor train representing $\frac{1}{\mathbf{r}}$ and the Laplace operator. After the construction of the corresponding Fock operator the HF equations are solved using the DMRG-algorithm with an increase of the total HF energy as stopping criterion. After this the orthogonality of the orbitals is recovered by a Gram-Schmidt orthogonalization. Subsequently the Fock matrix is constructed and diagonalized providing a rotated basis. If the HF energy is not considered converged, a new Fock operator is calculated and the process is repeated. After convergence the orbitals can be interpolated to a finer grid and solved there for a higher precision.}}
    \label{fig:minimization}
\end{figure}
A single Slater determinant and, by that, also its energy, is, like Eqs.~\eqref{eq:slater_determinant} and \eqref{eq:hf_energy} suggest, only dependent on the actual basis functions that are occupied by the electrons.
This is a very important point and can lead to confusion when introducing the basis set approximation, i.e., expressing the one-electron orbitals $\phi(\mathbf{r})$ by more basis functions than there are electrons in the system.
With this approximation the HF equations lead to the so called Roothaan-Hall equations and are solved by finding the optimal linear combinations for the occupied orbitals.

Our proposed minimization routine (see Fig.~\ref{fig:minimization}) is based on the QTT representation of the orbitals on a three-dimensional Cartesian grid.
In contrast to the aforementioned linear combination ansatz, the QTT ansatz uses only as many basis functions as there are electrons in the molecule.
Hence, the optimization parameters are not the coefficients, which describe linear combinations of predefined basis functions, but directly define the functional form of the occupied orbitals as entries in the tensor trains.

The minimization is driven by solving the pseudo-eigenvalue HF equations for each orbital $\phi'_a(\mathbf{r})$ separately using the DMRG algorithm with the Fock operator as Hamiltonian.
Being dependent on the one-electron wave functions $\phi_a(\mathbf{r})$, the operators need to be constructed either by an initial guess or by the results of the previous iteration.
As the improvement of a single eigenvalue, i.e., the orbital energy, does not necessarily lead to an improvement of the HF energy, a stopping criterion has to be defined.
We choose to stop the current iteration whenever there is an increase in energy of the determinant, which is built from the input orbitals and the orbital being optimized by the DMRG, i.e., we stop if
\begin{equation}
    E_{HF}(\phi_1, \dots, \phi'_a, \dots, \phi_{N_e}) > E_{HF}(\phi_1, \dots, \phi_a, \dots, \phi_{N_e}).
\end{equation}
This criterion is checked after every DMRG sweep.

After solving all $N_e$ pseudo-eigenvalue equations in this fashion, the new set of orbitals are not orthogonal anymore.
Hence, the orthogonality is restored by the Gram-Schmidt orthogonalization procedure.
The orbitals can now be used to calculate the Fock matrix, which is subsequently solved with a diagonalization.
By doing so, new linear combinations of the orbitals can be defined with the corresponding eigenvectors finalizing one iteration.
The minimization can be considered converged if the iterations do not lead to an improvement in the HF energy.

The calculations of accurate integrals with tensor trains depend on one hand on the quality of the approximation and on the other hand on the chosen grid spacing.
Usually more grid points lead to an increase in accuracy.
Since the Hartree-Fock method, as any other method in quantum chemistry, requires highly accurate integrals, a very fine grid seems to fit the minimization well.
That being said, fine grids have the drawback of being computationally intensive and can potentially lead to slower convergence.

To achieve the accuracy of a fine grid and the convergence as well as the speed of a course grid it is possible to minimize the tensor trains $\phi_a(\mathbf{r})$ with varying core numbers.
Starting with a low number of cores, encoding a courser grid, the proposed minimization routine can be used to optimize the orbitals.
After the minimization finishes, the tensor trains can be interpolated to a grid with 8-times as many values as before by interpolating each dimension ($x$, $y$, $z$) following section \ref{sec:Interpolation}.
This process of interpolating and optimizing can be repeated until the grid distances are small enough to obtain a desired accuracy.

\section{Numerical Experiments}
\label{sec:results}
In the first part of this section we describe and discuss the parameters that influence the precision of the one- and two-electron integrals and evaluate their optimal values for our minimization routine.
The second part focuses on the minimization routine itself and its application to fermionic many-particle systems with up to $10$ electrons (\ce{CH4}).

The numerical experiments were performed in Python and Julia.
In Python, the input depicted in Fig.~\ref{fig:minimization} was calculated by the cross interpolation algorithm implemented in TnTorch~\cite{UBS:22}, which is based on Ref.~\cite{Oseledets2010}.
All standard quantum chemical calculations were performed with PySCF~\cite{Sun2020}.
Julia and especially the package ITensor~\cite{ITensor} were used for all tensor train and tensor train matrix operations including the DMRG optimizations.

All calculations build upon some common features.
The operators $\frac{1}{|\mathbf{r}|}$, $\frac{1}{|\mathbf{r}_1-\mathbf{r}_2|}$ as well as the Laplace operator were the same for all calculations and were not truncated.
For the Laplace operator an accuracy of $8$ was applied, c.f. Tab.~\ref{tab:finite_diff_coeffs}.
The three-dimensional grid, on which the orbitals are defined, is chosen such that all dimensions have the same uniform distribution from $-15$ to $15$ Bohr radii according to Eq.~\eqref{eq:binary_fraction}.
Since we vary the precision $p_d$ from $13$ to $20$, the orbitals and the function $\frac{1}{|\mathbf{r}|}$ are approximated once by the TCI algorithm for $p_d=20$ and $p_d=21$ respectively and then interpolated to courser grids.
This eliminates any influence of the TCI algorithm with respect to different values of $p_d$ in our analysis.

\subsection{Precision of One- and Two-Electron Integrals}
Every quantum chemical simulation can be only as good as their corresponding building blocks, the basis functions as well as the operators.
Within the FDM-QTT picture neither the basis functions nor the operators are exact, so the correctness of this framework for the given problem has to be verified.
A simple way for performing such verification is the calculation of the one- and two-body integrals with a known method and a subsequent comparison to the fully-numerical ansatz described above.

Since we are concerned with representing molecular orbitals as tensor trains instead of linear combinations of predefined atomic orbitals, the starting point of each comparison is a converged LCAO-Hartree-Fock calculation using a Gaussian-type basis set.
The LCAO-MOs can be converted to QTT-MOs using cross interpolation by sampling the values of the Gaussian basis set in real space.
As can be seen in Eq.~\eqref{eq:hf_operators} there are four different contributions to the overall energy of a Slater determinant,
the kinetic energy, the potential energy, the Coulomb energy and the exchange energy.
The influence of the approximation can be therefore investigated separately for each contribution.

\begin{figure}[t!]
    \centering
    \includegraphics[width=0.90\linewidth]{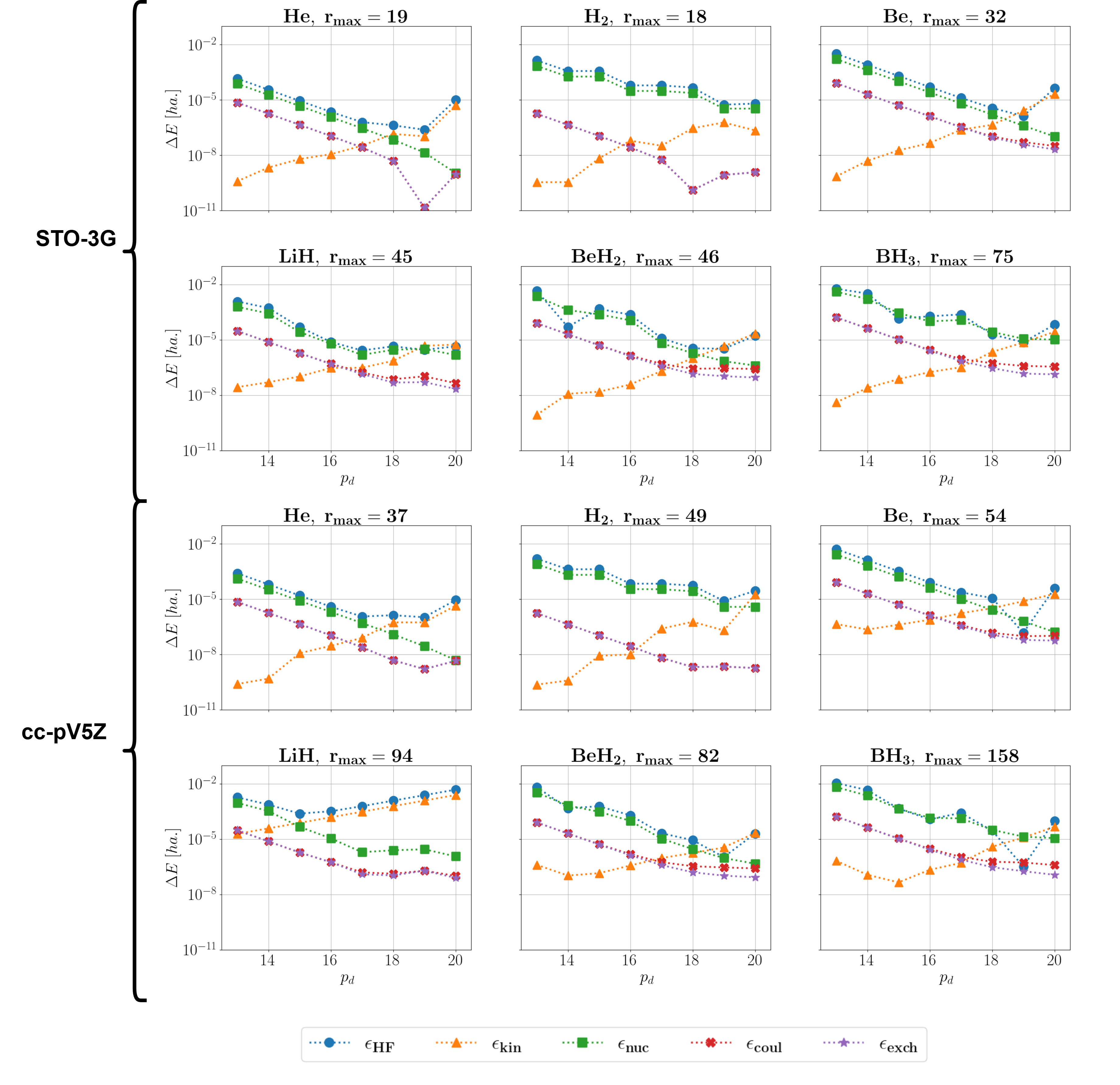}
    \caption{Error plots for the comparison of the LCAO- and QTT-ansatz. Depicted are the errors of the different contributions for the gaussian basis sets STO-3G and cc-pV5Z in dependence of $p_d$. {The errors are calculated according to~\eqref{eq:contr_err} with the orbital product approach}. The truncation cutoff was set to $10^{-16}$ for all orbitals. $r_{\text{max}}$ is the maximal rank that the orbitals of a corresponding system have and does not depend on $p_d$.}
    \label{fig:integrals}
\end{figure}

In a first step we show the approximation errors without the explicit computation of the Coulomb and exchange operators ({dubbed orbital product approach in the following}) for different values of $p_d$.
To get the corresponding contributions one can simply compress the terms $\phi_i\phi_j$ as tensor trains and perform the inner product with $g_{12}$.
In all our calculations $\phi_i\phi_j$ did not have larger ranks than the summed ranks of $\phi_i$ and $\phi_j$ showing that the product of the orbitals have an efficient QTT representation.
Fig.~\ref{fig:integrals} depicts the results for a set of selected even-electron systems and two different basis sets, STO-3G and cc-pV5Z respectively.
For both basis sets a truncation cutoff of $10^{-16}$ is applied to all orbitals.
The error of each contribution is calculated as the sum of the absolute differences between the QTT intagral and the LCAO integral for all orbitals:
{
\begin{equation}
\label{eq:contr_err}
\begin{aligned}
    \epsilon_{\text{kin}} &= \sum_i^{N_e} | \bra{\phi_i} -\frac{1}{2}\Delta \ket{\phi_i}_{\text{QTT}} - \bra{\phi_i} -\frac{1}{2}\Delta \ket{\phi_i}_{\text{LCAO}}|, \\
    \epsilon_{\text{nuc}} &= \sum_i^{N_e} | \bra{\phi_i} \sum_j^{N_n}\frac{Z_n}{|\mathbf{r}-\mathbf{R}_n|} \ket{\phi_i}_{\text{QTT}} - \bra{\phi_i} \sum_j^{N_n}\frac{Z_n}{|\mathbf{r}-\mathbf{R}_n|} \ket{\phi_i}_{\text{LCAO}}|, \\
    \epsilon_{\text{coul}} &= \sum_{i,j}^{N_e} | \bra{\phi_i(1)\phi_i(1)} \frac{1}{|\mathbf{r}_1-\mathbf{r}_2|} \ket{\phi_j(2)\phi_j(2)}_{\text{QTT}} - \bra{\phi_i(1)\phi_i(1)} \frac{1}{|\mathbf{r}_1-\mathbf{r}_2|} \ket{\phi_j(2)\phi_j(2)}_{\text{LCAO}}|, \\
    \epsilon_{\text{exch}} &= \sum_{i,j}^{N_e} | \bra{\phi_i(1)\phi_j(1)} \frac{1}{|\mathbf{r}_1-\mathbf{r}_2|} \ket{\phi_j(2)\phi_i(2)}_{\text{QTT}} - \bra{\phi_i(1)\phi_j(1)} \frac{1}{|\mathbf{r}_1-\mathbf{r}_2|} \ket{\phi_j(2)\phi_i(2)}_{\text{LCAO}}|.
\end{aligned}
\end{equation}
Due to cancellations these errors can be bigger than the error of the full Hartree-Fock energy.
The corresponding absolute energies of the reference LCAO calculations are listed in Tab.~\ref{tab:abs_lcao_energies}.}

\begin{table}[b!]
    \centering
    \begin{tabular}{|ccc|} \hline
        molecule   & STO-3G energy [ha.] & cc-pV5Z energy [ha.] \\ \hline
        \ce{He}    & -2.80778            & -2.86162             \\
        \ce{H_2}   & -1.87332            & -1.88875             \\
        \ce{Be}    & -14.35188           & -14.57301            \\
        \ce{LiH}   & -8.80946            & -8.93749             \\
        \ce{BeH_2} & -18.69852           & -18.91889            \\
        \ce{BH_3}  & -33.56634           & -33.89901            \\
        \ce{CH_4}  & -52.92503           & -53.41546            \\
        \hline
    \end{tabular}
    \caption{Absolute energy values of the converged LCAO-calculations.}
    \label{tab:abs_lcao_energies}
\end{table}

The first and most important observation is, that the Hartree-Fock energy converges for almost all investigated systems exponentially with increasing precision for $p_d \leq 18$, independent of the basis set.
Considering each contribution individually the nuclear, Coulomb and exchange energy show a similar exponential decay, with the nuclear energy being the most inaccurate.
The kinetic energy on the other hand seems to be very exact for low values of $p_d$, but gets exponentially worse for increasingly fine grid distances.
The reason for this is not the tensor train approximation itself but rather the errors introduced by the finite differences approach.
It is known that for grid distances smaller than $h=10^{-4}\approx 2^{-17}$ a higher precision than the standard double precision of $64$ bits is needed \cite{chertkov2016,Kazeev2017,Rakhuba2021}.
While the nuclear contribution bounds the error of the Hartree-Fock energy for some cases in the range $13 \leq p_d \leq 18$, it is the kinetic energy that becomes the main source of error for $p_d > 18$, while the Coulomb and exchange energies have no big effect on the error.

\begin{figure}
    \centering
    \includegraphics[width=0.9\linewidth]{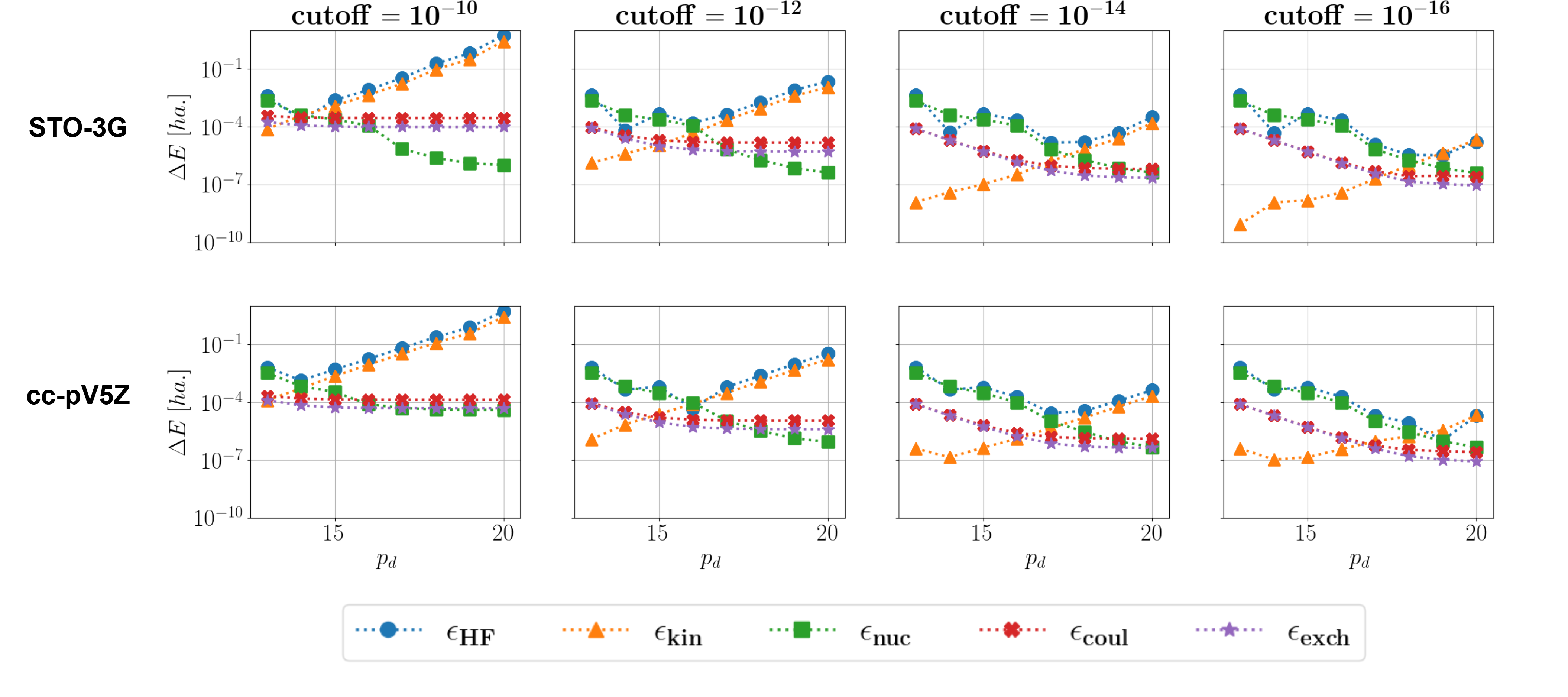}
    \caption{Dependency of the approximation error on the truncation cutoff for {\ce{BeH_2}}. The results were calculated according to~\eqref{eq:contr_err} {with the orbital product approach}.}
    \label{fig:truncation_error}
\end{figure}

\begin{figure}[b!]
    \centering
    \includegraphics[width=1.0\linewidth]{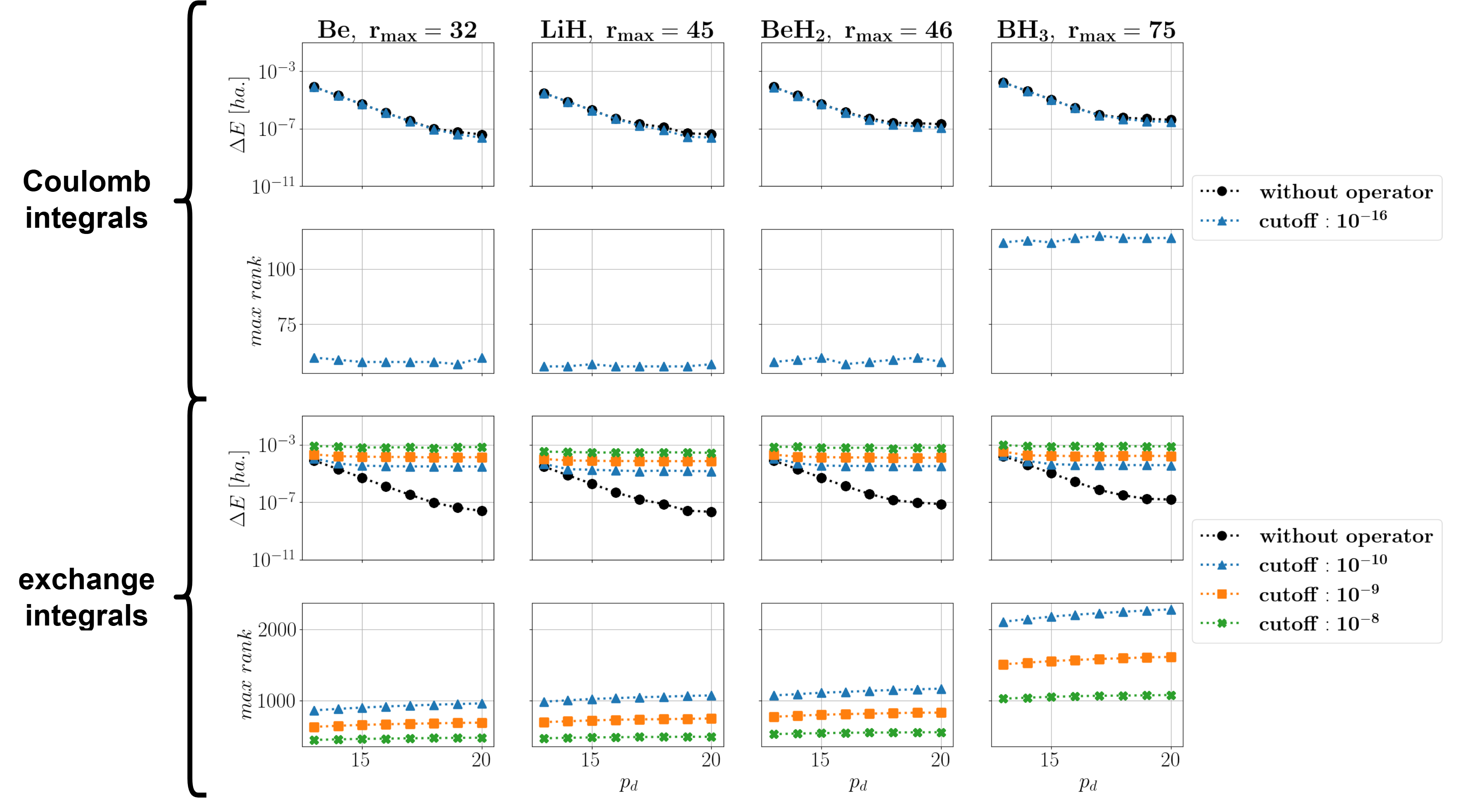}
    \caption{{Approximation error of the Coulomb and exchange integrals respectively, with and without (product approach) the explicit construction of the corresponding operators, J and K. The integrals can be calculated without the operators by building the orbital products $\phi_i\phi_j$ and using them subsequently in the inner products $\bra{\phi_i(1)\phi_i(1)}g_{12}\ket{\phi_j(2)\phi_j(2)}$ or $\bra{\phi_i(1)\phi_j(1)}g_{12}\ket{\phi_j(2)\phi_i(2)}$. The orbital products are approximated with a cutoff of $10^{-16}$. The error is calculated according to Eq.~\eqref{eq:contr_err}.}}
    \label{fig:operator_integrals_error}
\end{figure}

Having established that up to a certain degree, the QTT representation is capable of approximating the integrals needed in computational chemistry even for big basis sets, it is now time to turn towards the tuning of the truncation parameters.
The truncation parameters have a big influence on the accuracy of the one- and two electron-integrals and therefore on the success of solving the Hartree-Fock problem.
In Fig.~\ref{fig:truncation_error} the errors of the integrals for \ce{BeH_2} are depicted in dependency of the truncation cutoff with the same restrictions as stated above.
It can be clearly seen, that for low values of $p_d$ higher cutoffs are sufficient to describe the system accurately.
The finer the grid gets, the lower the cutoff has to be.
\ce{BeH_2} was chosen as an example, but the other atoms or molecules depicted in Fig.~\ref{fig:integrals} show a similar behavior.

A point that has not been addressed yet is the precision of the Coulomb and exchange operators.
For performing a DMRG calculations it is not possible to calculate the corresponding energy contributions by integrating the orbital product $\phi_i\phi_j$ with $g_{12}$.
Instead the operators $J$ and $K$ have to be explicitly created by the contractions depicted in Fig.~\ref{fig:operatorsl}.

Fig.~\ref{fig:operator_integrals_error} shows the error of the integrals $\bra{\phi_a}J_b\ket{\phi_a}$ and $\bra{\phi_a}K_b\ket{\phi_a}$ dependent on their truncation cutoff for the STO-3G basis set.
For the diagonal Coulomb operator the same cutoff as before, $10^{-16}$, can be applied and the energy matches with the orbital product approach.
In addition, the rank of the resulting operator $J_b$ seems to be independent of $p_d$ and behaves similar to the rank of $\phi_b\phi_b$.
The non-diagonal exchange operator $K_b$ has in comparison to $J_b$ large ranks, even for truncation cutoffs between $10^{-8}$ to $10^{-10}$.
For lower cutoffs the calculations become infeasible.
As can be expected for these large cutoffs, the error of the integrals does not show the exponential decay of the orbital product approach and stays constant in respect to finer grids.

The bad scaling of the exchange operators can be understood by considering the multiplication of $\phi_a(\mathbf{r})$ with $\phi_a(\mathbf{r}')$.
Unlike the Coulomb operator two orbitals are multiplied, that do not share the same electron coordinate.
This operation can be interpreted as an outer product of two tensor trains leading to a tensor train matrix, where one side represents $\mathbf{r}$ and the other one $\mathbf{r}'$.
The exploding ranks of $K$ suggest a poor approximation with this encoding.
Indeed, it can be shown that a tensor train with the alternating encoding where $\mathbf{r}$ and $\mathbf{r}'$ live on the same cores is not capable of representing $\phi_a(\mathbf{r})\phi_a(\mathbf{r}')$ with low ranks.
A block encoding separating $\mathbf{r}$ and $\mathbf{r}'$ on the other hand is able to describe the product efficiently.

\subsection{Hartree Fock for small atoms and molecules}
\begin{figure}[b!]
    \centering
    \includegraphics[width=1.00\linewidth]{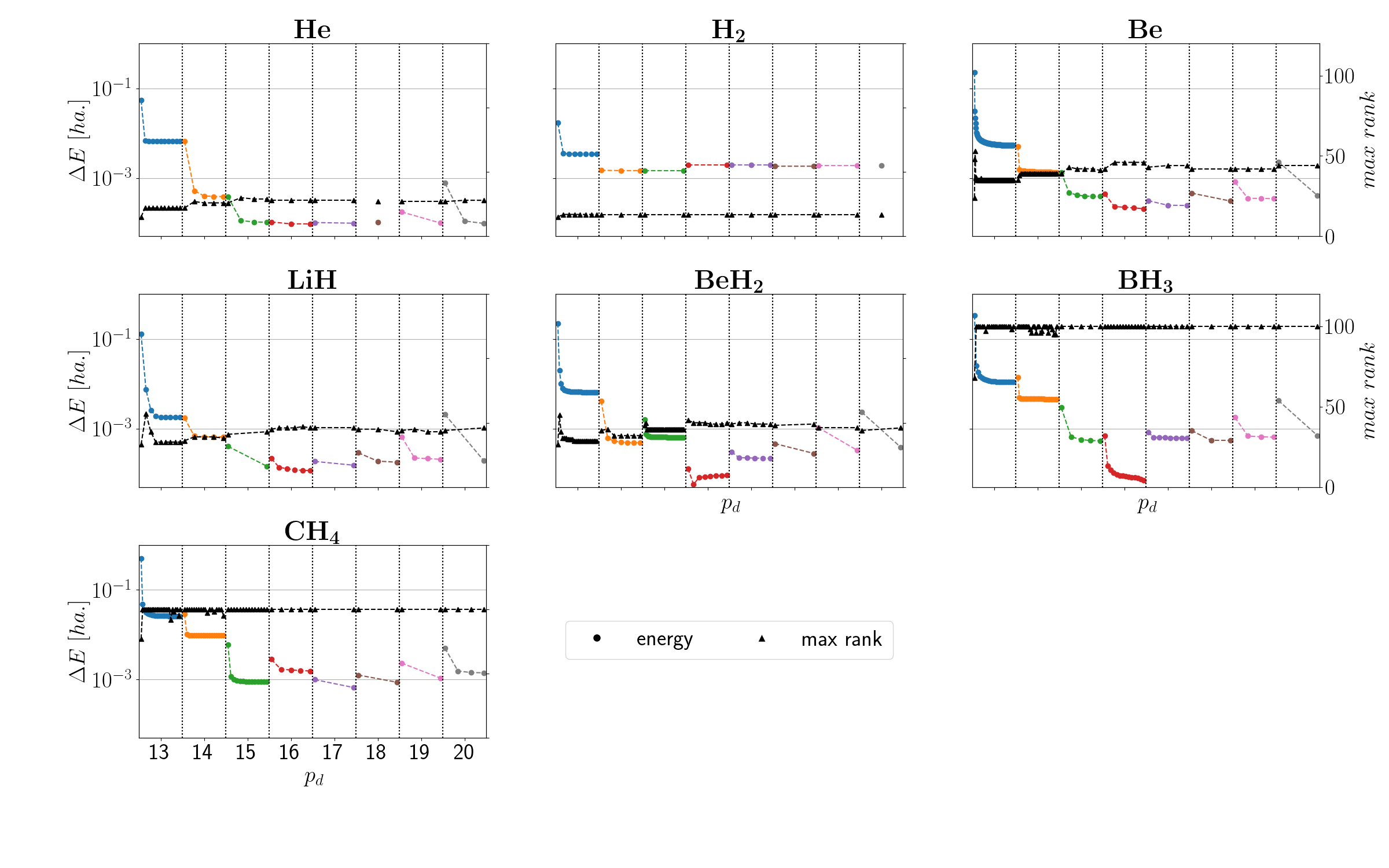}
    \caption{Minimization results for atomic and molecular systems with up to 10 electrons. Shown is the absolute energy difference of the QTT-FDM approach in respect to a LCAO calculation with the basis set cc-pV5Z for each step on grids from $p_d=13$ to $p_d=20$. Additionally the maximum rank of the orbitals is depicted. Each point corresponds to one iteration in the minimization routine.}
    \label{fig:optimization}
\end{figure}

After providing insights into the effect of the truncation parameter both on the integrals and the operators, the actual Hartree-Fock simulations can be discussed.
The simulations have been carried out according to Fig.~\ref{fig:minimization} starting with $p_d=13$.
After a calculation has been finished on a grid with $p_d$, the orbitals are interpolated onto the next one defined by $p_d+1$.
In this fashion the orbitals were optimized up to $p_d=20$.

Each time the grid was refined, the truncation cutoffs defining the approximation of the tensor train operations were adjusted.
Taking into account the results of the previous section, we defined two different cutoffs.
The first was applied during the creation of the exchange operator and remained constant in respect to $p_d$.
It was set to $10^{-9}$.
The second cutoff defined all remaining tensor train operations and was set in accordance with Fig.~\ref{fig:truncation_error} to values between $10^{-12}$ and $10^{-15}$.
Along the range $[13, 20]$ the cutoff is defined by $[10^{-12},10^{-13},10^{-14},10^{-15},10^{-15},10^{-15},10^{-15},10^{-15}]$.

The starting point of each optimization was a converged LCAO based Hartree-Fock calculation using the minimal basis set STO-3G.
As example for a big basis set with a near to optimum Hartree-Fock solution, the cc-pV5Z basis is taken to compare the QTT-results with.
Like before, the atoms and molecules chosen for the minimization were the even-electron atoms and hydrides with a electron count up to 8.
In addition to that the minimization routine was also tested for the 10-electron system Methane.
Due to computational restrictions the maximal rank of the orbitals and their products in \ce{BH_3} and \ce{CH_4} were set to 100. 

The optimization results are shown in Fig.~\ref{fig:optimization}.
Depicted are the absolute energy differences to the corresponding cc-pV5Z calculations as well as the maximum ranks of the orbitals in respect to $p_d$.
All systems except \ce{H_2} show a rapid convergence both on each grid individually and in respect to the grid distance up to $p_d\leq16$.
For larger values of $p_d$ the energy does not improve but rather stagnates or even gets worse.
Although the integrals shown in the previous section suggest a slightly larger range in which the Laplace operator is working correctly, it can be assumed that the kinetic energy prevents the minimization to converge more closely to the LCAO results.
Another reason can be found in the exchange operators only being calculated with a cutoff of $10^{-9}$.
As can be seen in Fig.~\ref{fig:operator_integrals_error} the energy difference of the exchange integrals calculated with the product approach on the one hand and the exchange operators on the other hand increases exponentially with $p_d$.

It can be concluded that, at least with the operators presented here, it is not possible to take full advantage of the exponential fine grids and by that of the exponential convergence of the nuclear and Coulomb integrals.
Nevertheless the results lie within an acceptable error range with respect to the LCAO calculations.
Especially \ce{BH_3} and \ce{CH_4} show a surprisingly good convergence even with the rank restriction to their orbitals.

\begin{figure}
    \centering
    \includegraphics[width=1.00\linewidth]{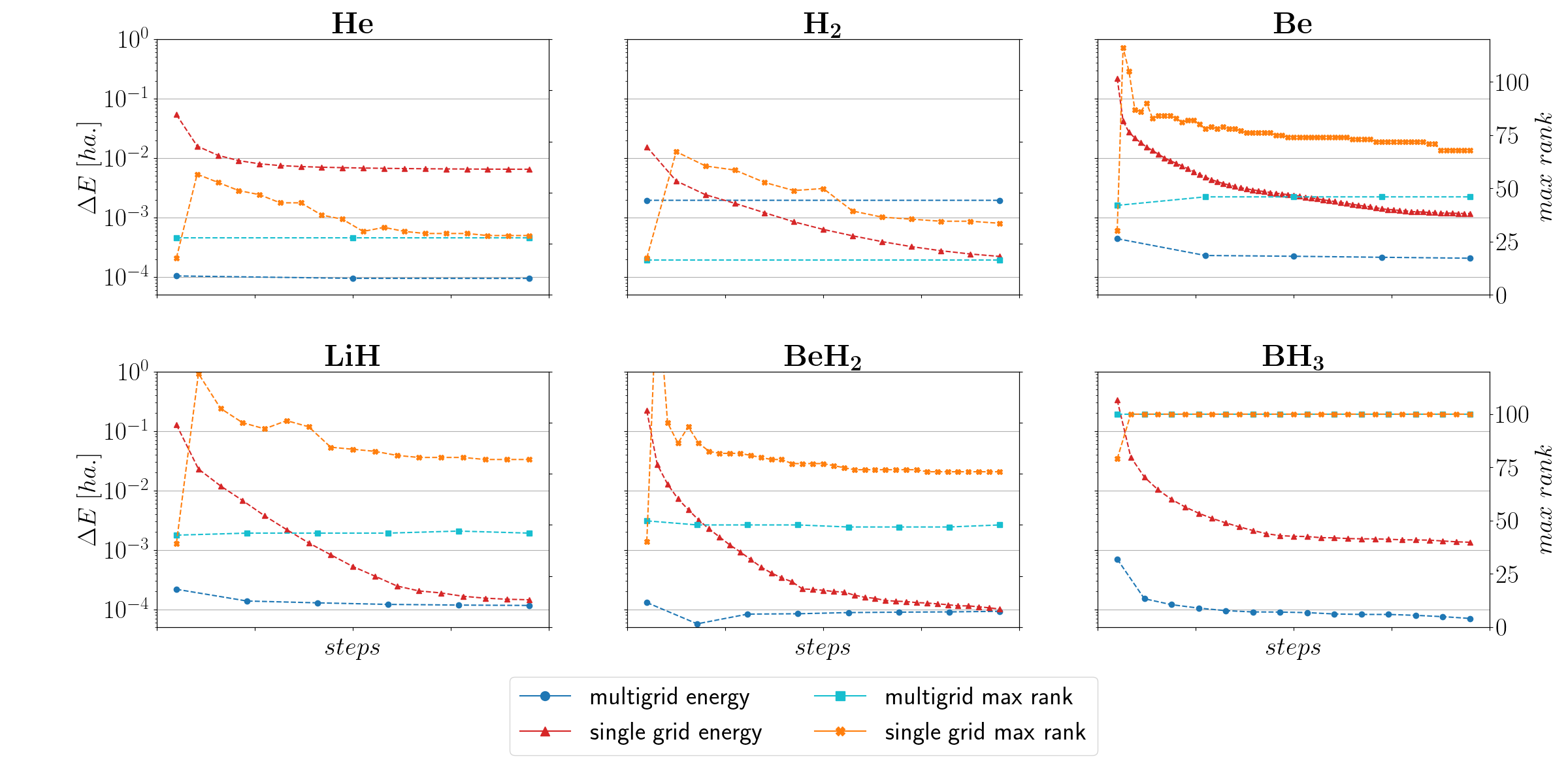}
    \caption{Comparison of the multi-grid and the single-grid approach for $p_d=16$. The multi-grid data was taken from the calculations depicted in Fig.~\ref{fig:optimization}. As before the energy difference in respect to a LCAO calculation with cc-pV5Z as basis is shown. Each point corresponds to one iteration in the minimization routine.}
    \label{fig:optimization_comparison}
\end{figure}

{As the next point,} we want to show the influence of optimizing the orbitals on multiple grids with increasing values of $p_d$ in respect to a single grid with a fixed $p_d$.
For the single grid calculations we chose $p_d=16$, since it can be assumed to be the point where the operators work best.
The comparison of both approaches is shown in Fig.~\ref{fig:optimization_comparison} {for all systems except \ce{CH_4} for which it was not possible to calculate the single grid solution due to slow convergence.}
Except for \ce{H_2} the minimizations performed with increasingly fine grids show smaller maximal ranks of their orbitals and a better or equally good energy.
The reason for the \ce{H_2} energy stagnating for the multi grid approach could be a local minima, which the DMRG algorithm cannot overcome.
\begin{figure}[b!]
    \centering
    \includegraphics[width=0.9\linewidth]{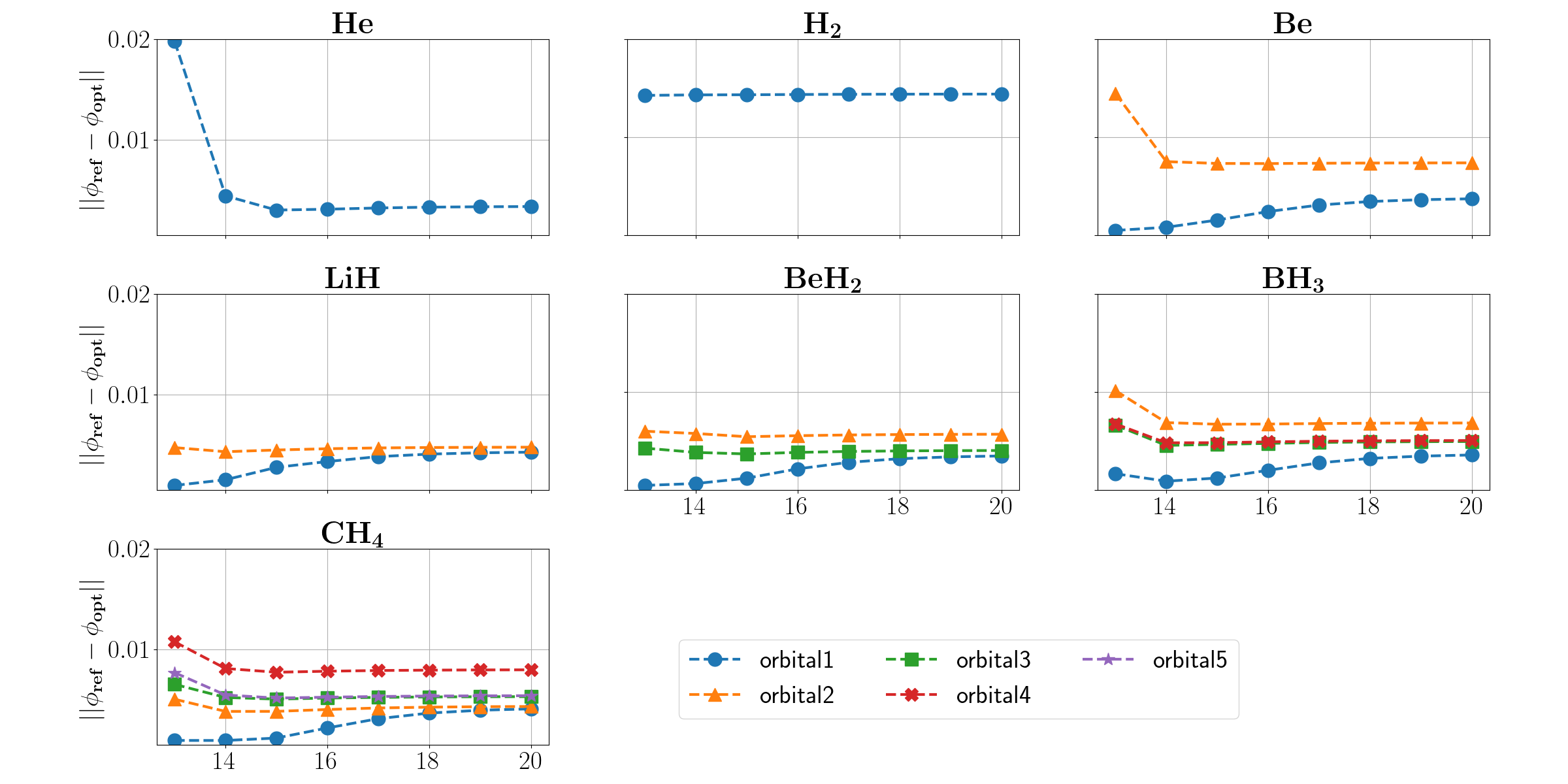}
    \caption{{Norms $||\phi^{ \text{LCAO}} - \phi^{ \text{QTT}}||$ of the converged QTT solutions and the cc-pV5Z reference in dependence of $p_d$ after a unitary rotation of $\phi_{\text{QTT}}$ to resemble $\phi_{\text{LCAO}}$ as closely as possible.}}
    \label{fig:norms}
\end{figure}

{
Since the same energy can be achieved with different wave functions, it is necessary to show that the QTT-FDM approach is not only reproducing the correct Hartree-Fock energy but also delivers a high-quality wave function.
The reference, which can be used to compare the QTT wave functions with, are the converged cc-pV5Z solutions.
Hartree-Fock wave functions are represented by Slater determinants, which depend on $3N_e$ spatial and $N_e$ spin coordinates, making them difficult to calculate and compare with each other.
Instead of the Slater determinant it is also possible to only consider the corresponding orbitals, which have a dimensionality of three.
If the orbitals are the same, the resulting Slater determinants will be as well.
To see if two sets of orbitals are equal, it is not enough to simply compare the individual orbitals ordered by their energy, since rotations and reflections among occupied orbitals lead to a change of these but do not to change the energy or the properties of the wave function itself.
}

{
Taking the tensor train quantics representation of the cc-pV5Z orbitals, which we calculate using the TCI algorithm, as a reference, the equivalence of the LCAO and QTT Slater determinants can be shown by expressing the cc-pV5Z orbitals as a linear combination of the QTT orbitals,
\begin{equation}
    \phi^{\text{LCAO}}_i(\mathbf{r}) = \sum_j^{N_e} c_{ij}\phi^{\text{QTT}}_j(\mathbf{r})
\end{equation}
with coefficients $c_{ij}=\braket{\phi^{\text{LCAO}}_i(\mathbf{r})|\phi^{\text{QTT}}_j(\mathbf{r})}$.
If the orbitals matched perfectly, the absolute value of the determinant of the matrix $c_{ij}$ would be $1$, leading to a rotation matrix and possible reflections of some orbitals.
Since the energy differences shown in Fig.~\ref{fig:optimization} indicate that the QTT-FDM approach did not fully achieve the precision of the cc-pV5Z wave function the absolute value of the determinants are not exactly $1$ but very close.
To preserve the minimization results and still resemble a rotation we perform the optimization of the coefficients $c'_{ij}$ subject to the function 
\begin{equation}
f(c'_{ij})=|c_{ij}-c'_{ij}|_F + ||\text{det}(c'_{ij})| - 1|,    
\end{equation}
where $|\cdot|_F$ denotes the Frobenius norm.
The resulting matrix has $|\det(c'_{ij})|=1$ and rotates the QTT orbitals as closely as possible to the reference while preserving the properties of the Slater determinant.
Using this rotation the orbitals of the minimization and the cc-pV5Z reference can be compared.
We calculate the quantics tensor train norm, 
\begin{equation}
||\phi^{ \text{LCAO}} - \phi^{ \text{QTT}}||\equiv \sqrt{\braket{\phi^{\text{LCAO}}_i(\mathbf{r})-\phi^{\text{QTT}}_j(\mathbf{r})|\phi^{\text{LCAO}}_i(\mathbf{r})-\phi^{\text{QTT}}_j(\mathbf{r})}}     
\end{equation}
to quantify the difference between both solutions, which can be used to bound the error on any observable (see Appendix~\ref{sec:appendix_b}, \ref{sec:appendix_c}).
Note, that this norm measures the distance between two orbitals on a \textit{discretized} space.
Indeed, for a given grid space $h=\frac{1}{2^p_d}$ one has
\begin{equation}
    ||\phi^{ \text{LCAO}} - \phi^{ \text{QTT}}|| = \sum_{i,j,k=1}^{2^p_d}|\phi(x_i,y_j,z_k)^{ \text{LCAO}} - \phi(x_i,y_j,z_k)^{ \text{QTT}}|.
\end{equation}
Thus, the norm does not say anything about the error due to the discretization.
The results are shown in Fig.~\ref{fig:norms}.
Except for the first orbital the norms decrease until $p_d=15$ and converge to a threshold which is roughly an order of magnitude smaller than the norms of the initial STO-3G orbitals (see Tab.~\ref{tab:norms}).
A possible explanation for this is that the norm of the first orbital, a s-orbital which describes the electrons close to the nuclei, is increasing due to the finer grid around the nuclei positions, a region that is hard to describe.
For \ce{H_2}, the difference between $\phi^{ \text{LCAO}}$ and $\phi^{ \text{QTT}}$ is rather high, underpinning the assumption that the minimization got stuck in a local minimum.
The small norms of the other molecules indicate a good description of the reference wave function showing that the QTT-FDM approach is not only able to reproduce LCAO energies but also their wave functions.
The rapidly converging norm with respect to $p_d$ reflect that the wave functions even at coarse grids lead to function values with a small difference compared to LCAO results.
Regarding the integration errors due to discretization one can expect that with finer grids this error can be suppressed exponentially with $1/2^p_d$, as can be seen in Fig.~\ref{fig:integrals}.
}
\begin{table}[t!]
    \centering
    \begin{tabular}{|c|ccccccc|} \hline
        molecule  & \ce{He}  & \ce{H_2} & \ce{Be} & \ce{LiH} & \ce{BeH_2} & \ce{BH_3} & \ce{CH_4}\\ \hline
        orbital 1 & 0.0587 & 0.0470 & 0.0262 & 0.0334 & 0.0253 & 0.0225 & 0.0209 \\
        orbital 2 &        &        & 0.1940 & 0.1230 & 0.0951 & 0.0458 & 0.0518 \\
        orbital 3 &        &        &        &        & 0.0845 & 0.0785 & 0.0755 \\
        orbital 4 &        &        &        &        &        & 0.0786 & 0.0755 \\
        orbital 5 &        &        &        &        &        &        & 0.0755 \\ \hline
    \end{tabular}
    \caption{{Norms of the converged LCAO results $||\phi_{\text{STO-3G}}-\phi_{\text{cc-pV5Z}}||$ calculated with $p_d=20$. Before the norm was calculated the STO-3G orbitals were subject to a unitary rotation which rotated them as closely as possible to the cc-pV5Z orbitals.}}
    \label{tab:norms}
\end{table}

\section{Conclusion and Outlook}
\label{sec:conclusion}
\begin{table}[b!]
    \renewcommand{\arraystretch}{2.0}
    \centering
    \begin{tabular}{|cccc|} \hline
         & rank dependency on $p_d \in [13, 20]$ & efficient QTT approximation & error source \\ \hline
         $\phi_i$& const. & yes & no \\
         $\phi_i\phi_j$& $\text{const.} < r_\text{max}(\phi_i) + r_\text{max}(\phi_j)$ & yes & no \\
         $\Delta$& $\text{const.} < 20$ & yes & yes \\
         $\frac{1}{|\mathbf{r}|}$& $\text{const.} \approx 170$ & yes & no \\
         $\frac{1}{|\mathbf{r}_1-\mathbf{r}_2|}$& $\text{const.} \approx 340$ & yes & no \\
         $\frac{\phi_i(\mathbf{r}_1)\phi_j(\mathbf{r}_1)}{|\mathbf{r}_1-\mathbf{r}_2|}$& $\text{const.} < r_\text{max}(\phi_i) + r_\text{max}(\phi_j)$ & yes & no \\
         $\frac{\phi_i(\mathbf{r}_1)\phi_j(\mathbf{r}_2)}{|\mathbf{r}_1-\mathbf{r}_2|}$& linear with slight slope & no & yes \\ \hline
    \end{tabular}
    \caption{Notes on the efficiency and accuracy of functions and operators needed to perform a mean-field calculation.}
    \label{tab:results_summary}
    \renewcommand{\arraystretch}{1}
\end{table}
In this paper we presented a proof-of-principle minimization routine to solve the Hartree-Fock problem within the finite-differences-method based on the quantized tensor train approximation.
The routine tackles the pseudo-eigenvalue equations with the DMRG algorithm on multiple, increasingly fine grids improving convergence.

A short summary of all relevant functions needed for performing the mean field calculations, as well as their properties in the QTT representation are given Tab.~\ref{tab:results_summary}.
The decomposition of orbitals as well as their product shows low ranks, a high accuracy and thus an efficient approximation.
Since the same is true for the inverse distance operators, it also holds for the Coulomb operator and thus the corresponding integrals show a with $p_d$ exponentially decaying reference solution.
Although the Laplace operator can be easily represented as a tensor train matrix, for grid distances smaller than $10^{-4}$ numerical instabilities arise, which lead to a low accuracy for the kinetic energy integrals.
This is a well known phenomenon and there are methods for preventing such behavior described in the literature~\cite{Rakhuba2021}.
In a future work, we plan to incorporate these methods such that simulations with arbitrary fine grids become feasible.
The only operator, which does not have an efficient representation in the QTT format, is the exchange operator.
It has large ranks even for large truncation cutoffs and leads to integrals with a low accuracy.
A possible solution for preventing the explicit construction of this operator could be based on its product structure in a custom DMRG algorithm.

The minimization routine was tested with even-electron systems with up to 10 electrons.
Almost all minimizations converged with respect to the LCAO approach with a large basis set and therefore support the validity of the QTT-FDM simulations.
Surprisingly, the rank restrictions for \ce{BH_3} and \ce{CH_4} did not lead to a worse convergence than the other calculations without restriction.
This circumstance indicates that also bigger molecules are in reach of the basic methodology.

The shown approach is not only valid for Hartree-Fock calculations but also for other mean-field methods like density functional theory.
Since the orbital products can be efficiently represented as tensor trains, so can the densities of Slater determinants.
Another interesting direction for further research would be the use of fully numerical basis functions for post Hartree-Fock calculations, which include orbital optimizations, like multi-configurational self-consistent field methods.
Because of the expressiveness of tensorized orbitals it can be expected that the number of basis functions needed for describing the electron correlation correctly could be reduced by a large factor.
This could make tensor trains an addition to deep neural networks like FermiNet~\cite{Pfau2020} or PauliNet~\cite{Hermann2020}, which were recently used to describe correlated molecular systems in first quantization.

\section*{Acknowledgements}
This work was supported by the research project \textit{Zentrum für Angewandtes Quantencomputing} (ZAQC), which is funded by the Hessian
Ministry for Digital Strategy and Innovation and the Hessian Ministry of Higher Education, Research and the Arts.
We thank Timon Scheiber for interesting and helpful discussions.

\begin{appendices}

\section{Band operators with variable encodings} 
\label{sec:appendix_a}
In this appendix, we introduce an algorithm that converts a band matrix into the tensor train matrix format for variable encodings.
A band matrix can be easily decomposed into a sum of matrices, where each matrix contains a single diagonal band, like
\begin{equation}
    \begin{pmatrix}
    0 & 1 & 0 & 0 \\
    2 & 0 & 1 & 0 \\
    0 & 2 & 0 & 1 \\
    0 & 0 & 2 & 0
    \end{pmatrix} = \begin{pmatrix}
    0 & 1 & 0 & 0 \\
    0 & 0 & 1 & 0 \\
    0 & 0 & 0 & 1 \\
    0 & 0 & 0 & 0
    \end{pmatrix} + \begin{pmatrix}
    0 & 0 & 0 & 0 \\
    2 & 0 & 0 & 0 \\
    0 & 2 & 0 & 0 \\
    0 & 0 & 2 & 0
    \end{pmatrix}.
    \label{eq:band_matrix_decomposition}
\end{equation}
Since the sum of two TTMs can be easily computed, the conversion task reduces to finding an algorithm, which is capable of creating a tensor train matrix describing a single diagonal band.
\begin{figure}[b!]
    \centering
    \includegraphics[width=1\linewidth]{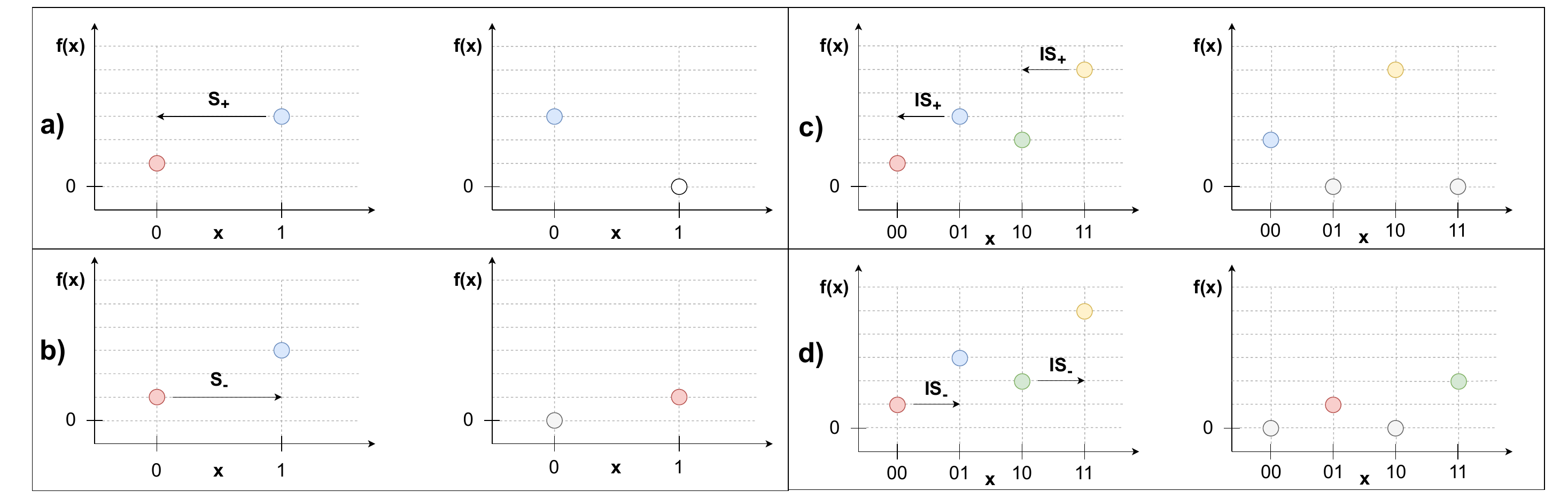}
    \caption{Effect of the $S_+$ and $S_-$ operators on two example systems with $p=1$ for a) and b) as well as with $p=2$ for c) and d).}
    \label{fig:bit_operator}
\end{figure}

From this point on we adopt some notation from spin-$1/2$ quantum physics.
Specifically, we need three $2\times 2$ matrices, which are the identity operator $I$ as well as the spin ladder operators $S_+$ and $S_-$:
\begin{equation}
    I = \begin{pmatrix}
    1 & 0 \\
    0 & 1
    \end{pmatrix},\ \ 
    S_+ = \begin{pmatrix}
    0 & 1 \\
    0 & 0
    \end{pmatrix},\ \ 
    S_- = \begin{pmatrix}
    0 & 0 \\
    1 & 0
    \end{pmatrix}.
    \label{eq:quantum_operators}
\end{equation}
Any rank-1 TTM can then be expressed by a sequence of operator strings defining the $2\times 2$ matrices of each core.
An example would be $IIIS_+$, which represents a TTM where the first three cores are defined by the identity matrix and the last core by the $S_+$ matrix.

Having established the notation the effect of the different operators on tensor trains can be described.
An operator string that is made of only identity operators is equivalent to the identity matrix and maps a tensor train therefore to itself.
Whenever an operator sequence contains the ladder operators $S_+$ or $S_-$ however the result is less trivial.
To understand the effect of such operators it is necessary to think of the tensor train as a whole, where it is not easily possible to address a single grid value, but every operation effects a multitude of grid values.
In the easiest case, in which the tensor train has only one core and is able to represents two values exactly, $S_+$ shifts the second value to the first one and sets the shifted value to zero (Fig.~\ref{fig:bit_operator} a).
In other words $S_+$ turns the value described with the bit parameter $b_{x,1}=1$ to the value with the bit parameter $b_{x,1}=0$.
$S_-$ performs the opposite mapping (Fig.~\ref{fig:bit_operator} b).

\begin{table}[t]
    \centering
    \begin{tabular}{c|c}
        shift as bitstrings & shift operator \\ \hline 
        0000 $\rightarrow$ 0001 & $IIIS_-$ \\
        0001 $\rightarrow$ 0010 & $IIS_-S_+$ \\
        0010 $\rightarrow$ 0011 & $IIIS_-$ \\
        0011 $\rightarrow$ 0100 & $IS_-S_+S_+$ \\
        0100 $\rightarrow$ 0101 & $IIIS_-$ \\
        0101 $\rightarrow$ 0110 & $IIS_-S_+$ \\
        0110 $\rightarrow$ 0111 & $IIIS_-$ \\
        0111 $\rightarrow$ 1000 & $S_-S_+S_+S_+$ \\
        1000 $\rightarrow$ 1001 & $IIIS_-$ \\
        1001 $\rightarrow$ 1010 & $IIS_-S_+$ \\
        ... & ...
    \end{tabular}
    \caption{Depiction how to find out the rank-1 TTMs for approximating a single band matrix with a positive distance of one for a function $f(x)$ and $p=4$.}
    \label{tab:shifting_procedure}
\end{table}

When considering a tensor train with two cores four values are in total approximated.
The operators now consist of two $2\times 2$ matrices like $IS_+$ or $IS-$ and effect not only two but all four values.
For example, $IS_+$ shifts the values defined by the bitstrings $01$ and $11$ to the values of $00$ and $10$ (Fig.~\ref{fig:bit_operator} c).
$IS_-$ on the contrary shifts the values defined by the bitstrings $00$ and $10$ to the values of $01$ and $11$ (Fig.~\ref{fig:bit_operator} d).
As a general rule the lowering operators $S_-$ at position $i$ turn the corresponding bit values from $1$ to $0$, while the raising operators $S_+$ turn the bit values from $0$ to $1$.

A single band matrix can be understood as a shifting operation where the distance of the band to the main diagonal is the number of grid points by which a vector is being shifted.
If the distance is positive, the vector is shifted to the right, while a negative distance indicates a shift to the left (see Fig.~\ref{fig:band_operator_shifting}).

\begin{figure}[t!]
    \centering
    \includegraphics[width=0.9\linewidth]{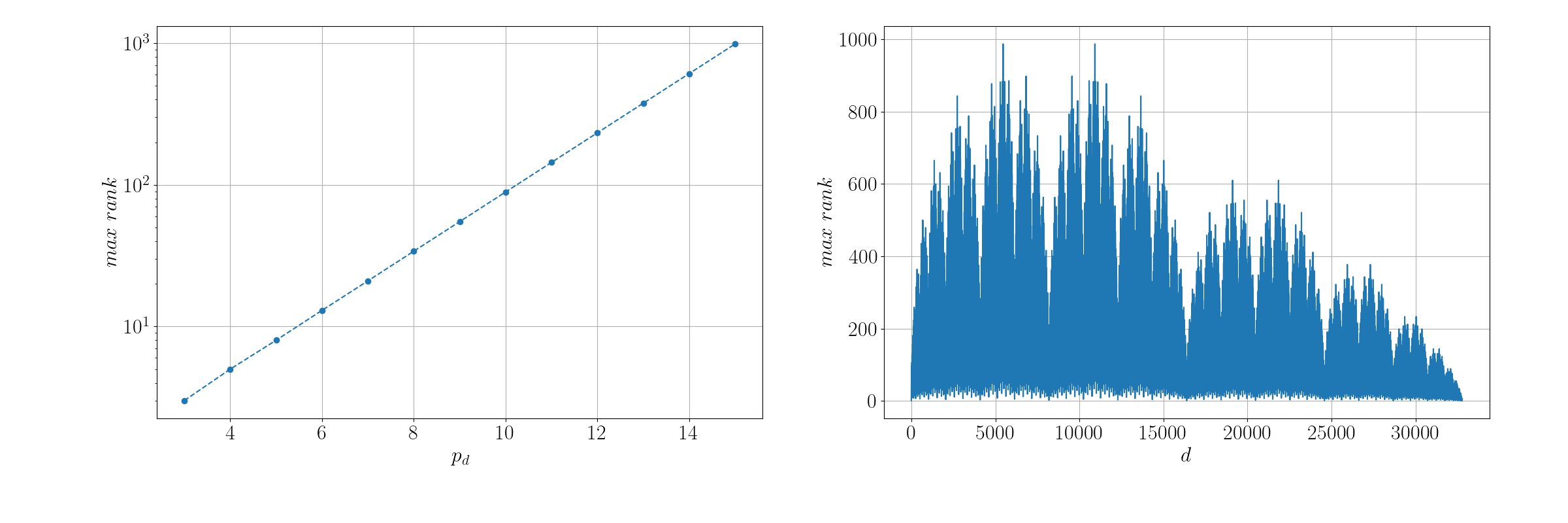}
    \caption{On the left side, the maximum rank of a TTM describing a single band operator in dependence of $p_d$ is depicted. On the right side, the rank for $p_d=15$ is plotted against the distance $d$ to the main diagonal.}
    \label{fig:band_ops_algo}
\end{figure}

\begin{algorithm}[b!]
\textbf{Input:} distance, $p$ \Comment{distance to the main diagonal and bit count}\\
\textbf{Output:} ops \Comment{operator sum defining the shift}
\newline
\begin{algorithmic}[1]
\Require $p > 0$ \Comment{bit count}
\Require $\text{distance} \neq 0$ \Comment{distance from the main diagonal}
\State $d = |\text{distance}|$
\State $\text{set} = \emptyset$ \Comment{empty set for tuples with two entries}
\For{$i = 0$ \textbf{to} $d - 1$}
    \For{$j = i$ \textbf{to} $2^{p} - 1$ \textbf{step} $d$}
        \State $v = j \oplus (j + d)$ \Comment{bitwise XOR $\oplus$}
        \State $\text{set} = \text{set} \cup \{(j + d) \land v, j \land v\}$ \Comment{bitwise AND $\land$}
    \EndFor
\EndFor

\State
\State $\text{ops}=\textbf{0}$
\For{term \textbf{in} $\text{set}$}
    \State $\text{op} = \textbf{I}\ \text{repeated}\ p\ \text{times}$ \Comment{example: $p=4$ leads to $\text{op}=IIII$}
    \For{$i=0$ \textbf{to} $p$}
        \If{$i\text{th-bit}$ \textbf{of} $\text{term}[0]$ \textbf{is set}}
            \State $\text{op}[i] = S_+$
        \ElsIf{$i\text{th-bit}$ \textbf{of} $\text{term}[1]$ \textbf{is set}} \Comment{bits of term[0] and term[1] cannot overlap}
            \State $\text{op}[i] = S_-$ \Comment{$\text{term}[0] \land \text{term}[1] = 0$}
        \EndIf
    \EndFor
    \If{$\text{distance} > 0$}
        \State $\text{ops} \mathrel{+}= \text{op}$
    \ElsIf{$\text{distance} < 0$}
        \State $\text{ops} \mathrel{+}= \text{op}^\dagger$ \Comment{example: $op=IIS_+S_+$, $op^\dagger=IIS_-S_-$}
    \EndIf
\EndFor
\end{algorithmic}
\caption{Algorithm for finding the operator sum of a single band diagonal.}
\label{alg:band_op_algorithm}
\end{algorithm}

To construct a tensor train matrix for this shifting operation, the first step is to describe every shift of each value by a specified distance individually with the operators defined in Eq.~\eqref{eq:quantum_operators}.
Tab.~\ref{tab:shifting_procedure} depicts the procedure for a small example.
As can be seen various operations are needed for describing the shift.
The reason for this behaviour can be understood by considering Fig.~\ref{fig:bit_operator}.
Every operator $IIS_+...$ shifts multiple values to new positions but assigns zero to the old positions.
The old positions can then be filled by shifting values with another operator.
Summing over all the operators without duplications leads to the same behaviour as a single band matrix and is therefore the solution of the stated problem.
The essential point why band operators can be efficiently represented by tensor train matrices lies within the fact that the operations shown in table~\ref{tab:shifting_procedure} start repeating themselves and therefore lead to a tensor train matrix with a rank much smaller then $2^{p_d}$.
Fig.~\ref{fig:band_ops_algo} shows the maximum rank of a single band operator in dependence of $p_d$ for a single dimension as well as the explicit dependency of the single band operator in respect to the distance from the main diagonal for $p_d=15$.
Although way less operators are needed then $2^{p_d}$ the scaling of the hardest to calculate distance is still exponential in respect to $p_d$.

Algorithm~\ref{alg:band_op_algorithm} describes the general process of computing the operator sum ($IIS_-+IS_-S_++...$) for a tensor train matrix, which represents a single band matrix where all entries have the value one.
Because the algorithm is based on shifting of every value of the QTT grid, the scaling is exponential.
This however is not a problem if one would try to calculate for example the laplace operator, since only band operators for single dimensions with distances $d\ll 2^{p_d}$ are needed.
As an example, consider a tensor train defined on a Cartesian grid $f(x,y,z)$ where $p_x=p_y=p_z=20$.
The grid describes $2^{60}\approx10^{18}$ points, so calculating a full band operator with the above developed algorithm is infeasible.
For a single dimension the cost reduces to $2^{20}\approx10^6$, which can be easily calculated.

\section{Influence of small wave function differences on observables}
\label{sec:appendix_b}
{
The difference of two wave functions, $\Psi$ and $\Phi$, which describe the same physical system, is directly connected to the difference of their observables and can be expressed by the inequality
\begin{equation}
\label{eq:obs_inequality}
    |\bra{\Psi}\hat{O}\ket{\Psi} - \bra{\Phi}\hat{O}\ket{\Phi}| \leq 2||\hat{O}|| \cdot ||\Phi-\Psi||.
\end{equation}
where $\hat{O}$ is an hermitian operator of the observable and $||.||$ denotes the norm of a wave function or an operator.
The deduction of this expression is based upon the triangular inequality
\begin{equation}
    ||\mathbf{u}+\mathbf{v}|| \leq ||\mathbf{u}|| + ||\mathbf{v}||
\end{equation}
and the Cauchy-Schwartz inequality
\begin{equation}
    |\braket{\mathbf{u}|\mathbf{v}}| \leq ||\mathbf{u}|| \cdot ||\mathbf{v}||
\end{equation}
with $\mathbf{u}$ and $\mathbf{v}$ as vectors in a Hilbert space.
Defining the difference of the wave functions as $\Delta=\Phi-\Psi$ we start with
\begin{equation}
    \bra{\Phi}\hat{O}\ket{\Phi} = \bra{\Psi+\Delta}\hat{O}\ket{\Psi+\Delta}.
\end{equation}
Expanding the terms and a subtraction by $\bra{\Psi}\hat{O}\ket{\Psi}$ leads to 
\begin{equation}
    |\bra{\Phi}\hat{O}\ket{\Phi} - \bra{\Psi}\hat{O}\ket{\Psi}| = |\bra{\Psi}\hat{O}\ket{\Delta} + \bra{\Delta}\hat{O}\ket{\Psi} + \bra{\Delta}\hat{O}\ket{\Delta}|,
\end{equation}
which, using the triangular inequality, leads to
\begin{equation}
    |\bra{\Phi}\hat{O}\ket{\Phi} - \bra{\Psi}\hat{O}\ket{\Psi}| \leq |\bra{\Psi}\hat{O}\ket{\Delta}| + |\bra{\Delta}\hat{O}\ket{\Psi}| + |\bra{\Delta}\hat{O}\ket{\Delta}|=    2 |\bra{\Psi}\hat{O}\ket{\Delta}| + |\bra{\Delta}\hat{O}\ket{\Delta}|.
\end{equation}
Using the Cauchy-Schwartz inequality and the assumption that $\Psi$ and $\Phi$ are normalized, the terms can be transformed to
\begin{equation}
\begin{aligned}
    |\bra{\Psi}\hat{O}\ket{\Delta}| &\leq ||\hat{O}\Delta||,\\
    |\bra{\Delta}\hat{O}\ket{\Delta}| &\leq ||\Delta|| \cdot ||\hat{O}\ket{\Delta}||.
\end{aligned}
\end{equation}
Since the operator norm satisfies $||\hat{O}\ket{\Delta}|| = ||\hat{O}|| \cdot ||\Delta||$, the final inequality becomes
\begin{equation}
    |\bra{\Psi}\hat{O}\ket{\Psi} - \bra{\Phi}\hat{O}\ket{\Phi}| \leq ||\hat{O}|| \cdot (2||\Phi-\Psi|| + ||\Phi-\Psi||^2).
\end{equation}
To leading order in $||\Delta||=||\Phi-\Psi||$  we arrive at
\begin{equation}
    |\bra{\Psi}\hat{O}\ket{\Psi} - \bra{\Phi}\hat{O}\ket{\Phi}| \leq 2||\hat{O}|| \cdot ||\Phi-\Psi||.
\end{equation}
}

\section{Observables of Slater determinants}
\label{sec:appendix_c}
{
If the observable of a Slater determinant needs to be calculated it is sufficient to apply the operator to its orbitals $\phi$, like
\begin{equation}
    \braket{\hat{O}} = \sum_{i}^{N_e} \bra{\phi_i(\mathbf{r}_1)}\hat{O}_1(\mathbf{r}_1)\ket{\phi_i(\mathbf{r}_1)} + \sum_{ij}^{N_e} \bra{\phi_i(\mathbf{r}_1)\phi_j(\mathbf{r}_2)}\hat{O}_2(\mathbf{r}_1,\mathbf{r}_2)\ket{\phi_k(\mathbf{r}_1)\phi_l(\mathbf{r}_2)} + ... ,
\end{equation}
with $\hat{O}=\sum_i^N \hat{O}_i(...\mathbf{r}_i)$ as an arbitrary operator effecting $N$ electrons.
Combining this with \eqref{eq:obs_inequality} the difference in an observable of two slater determinants, $\Psi$ and $\Psi'$, which are based on slightly different orbitals can be quantified by
\begin{equation}
\begin{aligned}
    \bra{\Psi}\hat{O}\ket{\Psi} - \bra{\Psi'}\hat{O}\ket{\Psi'} =& \sum_{i}^{N_e} \bra{\phi_i}\hat{O}_1\ket{\phi} - \bra{\phi'_i}\hat{O}_1\ket{\phi'_i} +\\
    &\sum_{ij}^{N_e} \bra{\phi_i\phi_j}\hat{O}_2\ket{\phi_k\phi_l} - \bra{\phi'_i\phi'_j}\hat{O}_2\ket{\phi'_k\phi'_l} + ...\\
    \leq& \sum_i^{N_e} 2||\hat{O}_1|| \cdot ||\phi_i-\phi_i'|| + \sum_{ij}^{N_e} 2||\hat{O}_2|| \cdot ||\phi_i\phi_j-\phi_i'\phi_j'|| + ...
\end{aligned}
\end{equation}
The higher order norms $||\prod_i^N\phi_i(\mathbf{r}_i)-\prod_i^N\phi_i'(\mathbf{r}_i)||$ are bounded by the first order norm $||\phi_i(\mathbf{r})-\phi_i'(\mathbf{r})||$ via
\begin{equation}
    ||\prod_i^N\phi_i(\mathbf{r}_i)-\prod_i^N\phi_i'(\mathbf{r}_i)|| \leq \sum_i^N ||\phi_i(\mathbf{r}_i)-\phi_i'(\mathbf{r}_i)||.
\end{equation}
This equation can be derived by considering
\begin{equation}
    \prod_i^N\phi_i(\mathbf{r}_i)-\prod_i^N\phi_i'(\mathbf{r}_i) =
    \sum_i^N \left(\prod_j^{i-1}\phi_j(\mathbf{r}_i)\right)
    (\phi_i(\mathbf{r}_i)-\phi_i'(\mathbf{r}_i)) 
    \left(\prod_{j=i+1}^N\phi_j'(\mathbf{r}_j)\right).
\end{equation}
Taking the norm of each side and using that $||\phi_i||=1$ the equation becomes
\begin{equation}
    ||\prod_i^N\phi_i(\mathbf{r}_i)-\prod_i^N\phi_i'(\mathbf{r}_i)|| =
    \sum_i^N ||\phi_i(\mathbf{r}_i)-\phi_i'(\mathbf{r}_i)||.
\end{equation}
Together with the triangle inequality the above stated inequality can be shown.
}
\end{appendices}
\clearpage

\bibliographystyle{ieeetr}
\bibliography{main}

\begin{thebibliography}{10}

\bibitem{Jensen2017}
F.~Jensen, {\em Introduction to Computational Chemistry}.
\newblock New York Academy of Sciences Series, Newark: John Wiley \& Sons,
  Incorporated, 1st ed.~ed., 2017.
\newblock Description based on publisher supplied metadata and other sources.

\bibitem{White1989}
S.~R. White, J.~W. Wilkins, and M.~P. Teter, ``Finite-element method for
  electronic structure,'' {\em Physical Review B}, vol.~39, pp.~5819--5833,
  Mar. 1989.

\bibitem{Harrison2004}
R.~J. Harrison, G.~I. Fann, T.~Yanai, Z.~Gan, and G.~Beylkin, ``Multiresolution
  quantum chemistry: Basic theory and initial applications,'' {\em The Journal
  of Chemical Physics}, vol.~121, pp.~11587--11598, Dec. 2004.

\bibitem{Lehtola2019}
S.~Lehtola, ``A review on non‐relativistic, fully numerical electronic
  structure calculations on atoms and diatomic molecules,'' {\em International
  Journal of Quantum Chemistry}, vol.~119, May 2019.

\bibitem{Kobus2013}
J.~Kobus, ``A finite difference hartree–fock program for atoms and diatomic
  molecules,'' {\em Computer Physics Communications}, vol.~184, pp.~799--811,
  Mar. 2013.

\bibitem{Khoromskij2011a}
B.~N. Khoromskij, V.~Khoromskaia, and H.-J. Flad, ``Numerical solution of the
  hartree–fock equation in multilevel tensor-structured format,'' {\em SIAM
  Journal on Scientific Computing}, vol.~33, pp.~45--65, Jan. 2011.

\bibitem{Khoromskaia2018}
V.~Khoromskaia and B.~N. Khoromskij, {\em Tensor Numerical Methods in Quantum
  Chemistry}.
\newblock De Gruyter, June 2018.

\bibitem{jolly2024}
N.~Jolly, Y.~N. Fernández, and X.~Waintal, ``Tensorized orbitals for
  computational chemistry,'' {\em arXiv preprint arXiv:2308.03508}, 2024.

\bibitem{Marcati2022}
C.~Marcati, M.~Rakhuba, and C.~Schwab, ``Tensor rank bounds for point
  singularities in ℝ3,'' {\em Advances in Computational Mathematics},
  vol.~48, Apr. 2022.

\bibitem{Rakhuba2016}
M.~Rakhuba and I.~Oseledets, ``Grid-based electronic structure calculations:
  The tensor decomposition approach,'' {\em Journal of Computational Physics},
  vol.~312, pp.~19--30, May 2016.

\bibitem{Gourianov2022}
N.~Gourianov, M.~Lubasch, S.~Dolgov, Q.~Y. van~den Berg, H.~Babaee, P.~Givi,
  M.~Kiffner, and D.~Jaksch, ``A quantum-inspired approach to exploit
  turbulence structures,'' {\em Nature Computational Science}, vol.~2,
  pp.~30--37, Jan. 2022.

\bibitem{Gourianov2025}
N.~Gourianov, P.~Givi, D.~Jaksch, and S.~B. Pope, ``Tensor networks enable the
  calculation of turbulence probability distributions,'' {\em Science
  Advances}, vol.~11, Jan. 2025.

\bibitem{Dey2023}
D.~Dey, A.~Parvej, S.~Das, S.~K. Saha, M.~Kumar, S.~Ramasesha, and Z.~G. Soos,
  ``Density matrix renormalization group (dmrg) for interacting spin chains and
  ladders,'' {\em Journal of Chemical Sciences}, vol.~135, Mar. 2023.

\bibitem{Hastings2007}
M.~B. Hastings, ``An area law for one-dimensional quantum systems,'' {\em
  Journal of Statistical Mechanics: Theory and Experiment}, vol.~2007,
  pp.~P08024--P08024, Aug. 2007.

\bibitem{Ye2022}
E.~Ye and N.~F.~G. Loureiro, ``Quantum-inspired method for solving the
  vlasov-poisson equations,'' {\em Physical Review E}, vol.~106, p.~035208,
  Sept. 2022.

\bibitem{Paeckel2019}
S.~Paeckel, T.~Köhler, A.~Swoboda, S.~R. Manmana, U.~Schollwöck, and
  C.~Hubig, ``Time-evolution methods for matrix-product states,'' {\em Annals
  of Physics}, vol.~411, p.~167998, Dec. 2019.

\bibitem{Oseledets2009}
I.~V. Oseledets, ``Approximation of matrices with logarithmic number of
  parameters,'' {\em Doklady Mathematics}, vol.~80, pp.~653--654, Oct. 2009.

\bibitem{Khoromskij2011}
B.~N. Khoromskij, ``O(dlog n)-quantics approximation of n-d tensors in
  high-dimensional numerical modeling,'' {\em Constructive Approximation},
  vol.~34, pp.~257--280, Apr. 2011.

\bibitem{Kazeev2017}
V.~Kazeev and C.~Schwab, ``Quantized tensor-structured finite elements for
  second-order elliptic pdes in two dimensions,'' {\em Numerische Mathematik},
  vol.~138, pp.~133--190, July 2017.

\bibitem{Oseledets2011}
I.~V. Oseledets, ``Tensor-train decomposition,'' {\em SIAM Journal on
  Scientific Computing}, vol.~33, pp.~2295--2317, Jan. 2011.

\bibitem{Oseledets2010}
I.~Oseledets and E.~Tyrtyshnikov, ``Tt-cross approximation for multidimensional
  arrays,'' {\em Linear Algebra and its Applications}, vol.~432, pp.~70--88,
  Jan. 2010.

\bibitem{Savostyanov2011}
D.~Savostyanov and I.~Oseledets, ``Fast adaptive interpolation of
  multi-dimensional arrays in tensor train format,'' in {\em The 2011
  International Workshop on Multidimensional (nD) Systems}, pp.~1--8, IEEE,
  Sept. 2011.

\bibitem{Savostyanov2014}
D.~V. Savostyanov, ``Quasioptimality of maximum-volume cross interpolation of
  tensors,'' {\em Linear Algebra and its Applications}, vol.~458, pp.~217--244,
  Oct. 2014.

\bibitem{Dolgov2020}
S.~Dolgov and D.~Savostyanov, ``Parallel cross interpolation for high-precision
  calculation of high-dimensional integrals,'' {\em Computer Physics
  Communications}, vol.~246, p.~106869, Jan. 2020.

\bibitem{NunezFernandez2022}
Y.~Núñez~Fernández, M.~Jeannin, P.~T. Dumitrescu, T.~Kloss, J.~Kaye,
  O.~Parcollet, and X.~Waintal, ``Learning feynman diagrams with tensor
  trains,'' {\em Physical Review X}, vol.~12, p.~041018, Nov. 2022.

\bibitem{Ritter2024}
M.~K. Ritter, Y.~Núñez~Fernández, M.~Wallerberger, J.~von Delft,
  H.~Shinaoka, and X.~Waintal, ``Quantics tensor cross interpolation for
  high-resolution parsimonious representations of multivariate functions,''
  {\em Physical Review Letters}, vol.~132, p.~056501, Jan. 2024.

\bibitem{Kazeev2013}
V.~A. Kazeev, B.~N. Khoromskij, and E.~E. Tyrtyshnikov, ``Multilevel toeplitz
  matrices generated by tensor-structured vectors and convolution with
  logarithmic complexity,'' {\em SIAM Journal on Scientific Computing},
  vol.~35, pp.~A1511--A1536, Jan. 2013.

\bibitem{Fornberg1988}
B.~Fornberg, ``Generation of finite difference formulas on arbitrarily spaced
  grids,'' {\em Mathematics of Computation}, vol.~51, no.~184, pp.~699--706,
  1988.

\bibitem{GarciaRipoll2021}
J.~J. García-Ripoll, ``Quantum-inspired algorithms for multivariate analysis:
  from interpolation to partial differential equations,'' {\em Quantum},
  vol.~5, p.~431, Apr. 2021.

\bibitem{UBS:22}
M.~Usvyatsov, R.~Ballester-Ripoll, and K.~Schindler, ``tntorch: Tensor network
  learning with {PyTorch},'' {\em Journal of Machine Learning Research},
  vol.~23, no.~208, pp.~1--6, 2022.

\bibitem{Sun2020}
Q.~Sun, X.~Zhang, S.~Banerjee, P.~Bao, M.~Barbry, N.~S. Blunt, N.~A. Bogdanov,
  G.~H. Booth, J.~Chen, Z.-H. Cui, J.~J. Eriksen, Y.~Gao, S.~Guo, J.~Hermann,
  M.~R. Hermes, K.~Koh, P.~Koval, S.~Lehtola, Z.~Li, J.~Liu, N.~Mardirossian,
  J.~D. McClain, M.~Motta, B.~Mussard, H.~Q. Pham, A.~Pulkin, W.~Purwanto,
  P.~J. Robinson, E.~Ronca, E.~R. Sayfutyarova, M.~Scheurer, H.~F. Schurkus,
  J.~E.~T. Smith, C.~Sun, S.-N. Sun, S.~Upadhyay, L.~K. Wagner, X.~Wang,
  A.~White, J.~D. Whitfield, M.~J. Williamson, S.~Wouters, J.~Yang, J.~M. Yu,
  T.~Zhu, T.~C. Berkelbach, S.~Sharma, A.~Y. Sokolov, and G.~K.-L. Chan,
  ``Recent developments in the pyscf program package,'' {\em The Journal of
  Chemical Physics}, vol.~153, July 2020.

\bibitem{ITensor}
M.~Fishman, S.~R. White, and E.~M. Stoudenmire, ``{The ITensor Software Library
  for Tensor Network Calculations},'' {\em SciPost Phys. Codebases}, p.~4,
  2022.

\bibitem{chertkov2016}
A.~V. Chertkov, I.~V. Oseledets, and M.~V. Rakhuba, ``Robust discretization in
  quantized tensor train format for elliptic problems in two dimensions,'' {\em
  arXiv preprint arXiv:1612.01166}, 2016.

\bibitem{Rakhuba2021}
M.~Rakhuba, ``Robust alternating direction implicit solver in quantized tensor
  formats for a three-dimensional elliptic pde,'' {\em SIAM Journal on
  Scientific Computing}, vol.~43, pp.~A800--A827, Jan. 2021.

\bibitem{Pfau2020}
D.~Pfau, J.~S. Spencer, A.~G. D.~G. Matthews, and W.~M.~C. Foulkes, ``Ab initio
  solution of the many-electron schrödinger equation with deep neural
  networks,'' {\em Physical Review Research}, vol.~2, p.~033429, Sept. 2020.

\bibitem{Hermann2020}
J.~Hermann, Z.~Schätzle, and F.~Noé, ``Deep-neural-network solution of the
  electronic schrödinger equation,'' {\em Nature Chemistry}, vol.~12,
  pp.~891--897, Sept. 2020.

\end{thebibliography}

\end{document}